# Markets Evolution After the Credit Crunch

## Marco Bianchetti

Market Risk Management, Intesa Sanpaolo,
Piazza Paolo Ferrari, 10, 20121 Milan, Italy,
marco.bianchetti[AT]intesasanpaolo.com

## Mattia Carlicchi

Market Risk Management, Intesa Sanpaolo,
Piazza Paolo Ferrari, 10, 20121 Milan, Italy,
mattia.carlicchi[AT]intesasanpaolo.com

19 December 2012

## Abstract

We review the main changes in the interbank market after the financial crisis started in August 2007. In particular, we focus on the fixed income market and we analyse the most relevant empirical evidences regarding the divergence of the existing basis between interbank rates with different tenor, such as Libor and OIS. We also discuss a qualitative explanation of these effects based on the consideration of credit and liquidity variables. Then, we focus our attention on the diffusion of collateral agreements among OTC derivatives market counterparties, and on the consequent change of paradigm for pricing derivatives. We illustrate the main qualitative features of the new market practice, called CSA discounting, and we point out the most relevant issues for market players associated to its adoption.

### Acknowledgements and disclaimer

A version of this paper appears as a chapter in "Interest Rate Modelling After the Financial Crisis", editors M. Bianchetti and M. Morini, Risk Books, 2013.
The authors gratefully acknowledge fruitful interactions with M. Morini and many colleagues at Intesa Sanpaolo and Banca IMI.
The views and the opinions expressed here are those of the author and do not represent the opinions of his employer. They are not responsible for any use that may be made of these contents. No part of this presentation is intended to influence investment decisions or promote any product or service.

# Table of Contents

# 1. Introduction

The financial crisis begun in the second half of 2007 has triggered, among many consequences, a deep evolution phase of the classical framework adopted for trading derivatives. In particular, credit and liquidity risks were found to have macroscopical impacts on the prices of financial instruments, both plain vanillas and exotics. The market has learnt the lesson and persistently shows such effects. These are clearly visible in the market quotes of plain vanilla interest rate instruments, such as Deposits, Forward Rate Agreements (FRA), Swaps (IRS) and options (European Caps, Floors and Swaptions). Since August 2007, the primary interest rates of the interbank market, e.g. Libor, Euribor, Eonia, and Federal Funds rate[1], display large basis spreads that have raised up to 200 basis points. Similar divergences are also found between FRA rates and the forward rates implied by two consecutive Deposits, and similarly, among Swap rates with different floating leg tenors (Basis Swaps).

After the financial crisis, the standard no-arbitrage framework adopted to price derivatives, developed over forty years following the Copernican Revolution of Black and Scholes (1973) and Merton (1973), became obsolete. Familiar relations described on standard textbooks (see e.g. Brigo and Mercurio 2006, Andersen and Piterbarg 2012, Hull 2008), such as the basic definition of forward interest rates, or the swap pricing formula, had to be abandoned. Also the fundamental idea of the construction of a single risk free yield curve, reflecting at the same time the present cost of funding of future cash flows and the level of forward rates, has been ruled out. The financial community has thus been forced to start the development of a new theoretical framework, including a larger set of relevant risk factors, and to review from scratch the no-arbitrage models used on the market for derivatives' pricing and risk analysis. A relevant feature of the post-crisis market is given by the consideration of collateral agreements in the pricing framework of OTC trades.

The paper is organized as follows. In section 2 we report the main changes and market evidences that characterize the most relevant interest rates of the interbank market since the explosion of the financial crisis. In particular, we focus on the EUR market and we analyze the relation between Euribor and Eonia market rates with different tenors, as observed in market quotations of Deposits, FRA, Swaps, Basis Swaps and Overnight Indexed Swaps (OIS). We argue that the financial crisis has sparked market liquidity and credit risk perception, that has been promptly reflected in the interest rates dynamics through increased and differentiated risk premia. We present a qualitative analysis of the Euribor – Eonia basis where we highlight the impacts of the credit and liquidity risk factor by introducing synthetic proxies that gauge the evolution of these two components during the period Jan. 2007 – Dec. 2011. In Section 3 we introduce the collateralization mechanics and the corresponding pricing methodology, called CSA discounting, that has been adopted by financial institutions to price collateralized trades, showing the consequences on market quotations of plain vanilla European Caps, Floors and Swaptions. Finally, we discuss the most relevant market issues regarding collateral management and pricing approach banks has to deal with while they fine tune their market practice and architecture to the evolved market framework. Conclusions are drawn in section 4.

# 2. The Interbank Market After the Financial Crisis

In this section we discuss the most relevant market data showing the main consequences of the financial crisis that started in August 2007. In particular, we focus our attention on Euribor and

---

[1] Libor, sponsored by the British Banking Association (BBA), is quoted in all the major currencies and is the reference rate for international Over-The-Counter (OTC) transactions (see www.bbalibor.com). Euribor and Eonia, sponsored by the European Banking Federation (EBF), are the reference rates for OTC transactions in the Euro market (see www.euribor.org). The Federal Funds rate is a primary rate of the USD market and is set by the Federal Open Market Committee (FOMC) accordingly to the monetary policy decisions of the Federal Reserve (FED) (see http://www.federalreserve.gov).

Eonia market rates, as observed in market quotations of standard plain vanilla interest rate linear instruments, such as Deposits, FRA, Swaps, Basis Swaps and OIS[2]. We analyse the basis spread among Euribor and Eonia rate with different tenors, which affects, directly or implicitly, comparable market quotations of Deposits, FRAs, Swaps, Basis Swaps and OIS. Similar results hold for other currencies, e.g. USD Libor and Federal Funds rates (see. e.g. Mercurio 2009, 2010).

Moreover, we report some market evidences trying to assess, in a qualitative way, the connections between credit and liquidity risk factors and the interbank market rates dynamics. To this aim, we consider quoted CDS spreads related to primary financial institutions of the EUR market and the volumes of the European Central Bank's monetary policy operations and balance sheet items during the crisis.

## 2.1. Euribor – Eonia Basis

One of the most relevant impacts of the financial turmoil over the interest rate market dynamics is the explosion of the basis between Euribor and Eonia rates. Before August 2007 these two rates displayed strictly overlapping trends, differing by no more than 6 basis points (bps). In August 2007 there has been a sudden increase of the Euribor rate and a simultaneous decrease of the OIS rate, that lead to the explosion of the corresponding basis spread.

The reason of the abrupt divergence of the Euribor-Eonia basis can be explained by considering both the impact of the crisis on the credit and liquidity risk perception of the market and the monetary policy decisions adopted by international authorities in response to the financial turmoil, coupled with the different financial meaning and dynamics of these rates.

- The Euribor rate is the reference rate for over-the-counter (OTC) transactions in the Euro area. It is defined as "the rate at which Euro interbank Deposits are being offered within the EMU zone by one prime bank to another at 11:00 a.m. Brussels time". The rate fixings for a strip of 15 maturities, ranging from one day to one year, are constructed as the trimmed average of the individual fixings (excluding the highest and lowest 15% tails) submitted by a panel of banks. The Contribution Panel is composed, as of September 2010, by 42 banks, selected among the EU banks with the highest business volume and credit standing in the Euro zone money markets, plus some large international bank from non-EU countries with important euro zone operations. Thus, Euribor rates reflect the average cost of funding of EU banks in the EUR interbank market at each given maturity.

- The Eonia rate is the reference rate for overnight OTC transactions in the Euro area. It is constructed as the average rate of the overnight transactions (one day maturity deposits) executed during a given business day by a panel of banks on the interbank money market, weighted by the corresponding transaction volumes. The Eonia Contribution Panel coincides with the Euribor Contribution Panel. Thus, Eonia rate includes information on the short term (overnight) liquidity expectations of banks in the Euro money market. It is also used by the European Central Bank (ECB) as a method of effecting and observing the transmission of its monetary policy actions. Furthermore, the daily tenor of the Eonia rate makes negligible the credit and liquidity risks reflected on it: for this reason the OIS rates are considered the best proxies available in the market for the risk-free rate.

---

[2] The Overnight Index Swap (OIS) is a swap with a fixed leg versus a floating leg indexed to the overnight rate (daily compounded over the coupon period). The Euro market quotes a standard OIS strip indexed to Eonia up to 30 years maturity. OISs with maturity up to 1 year settle a single coupon.

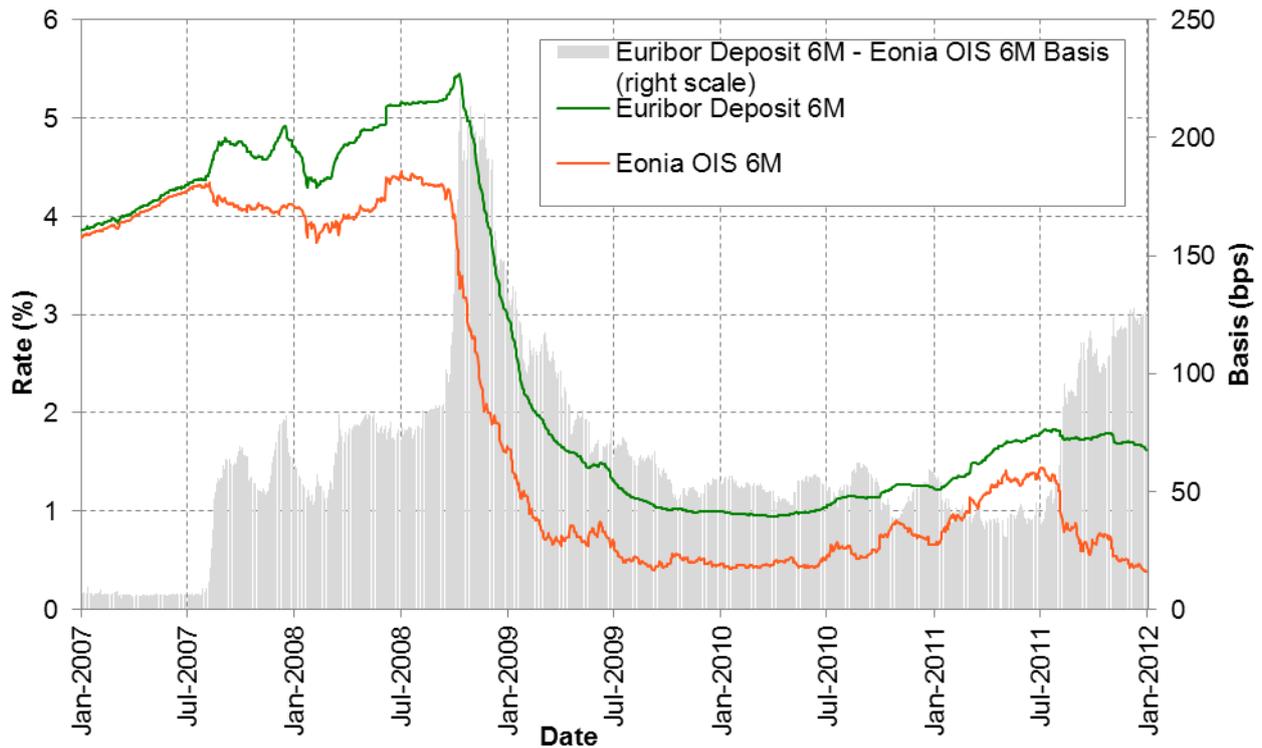

**Figure 1:** historical series of Euribor Deposit 6M rate versus Eonia OIS 6M rate. The corresponding spread is shown on the right scale (Jan. 2007 – Dec. 2011 window, source: Bloomberg).

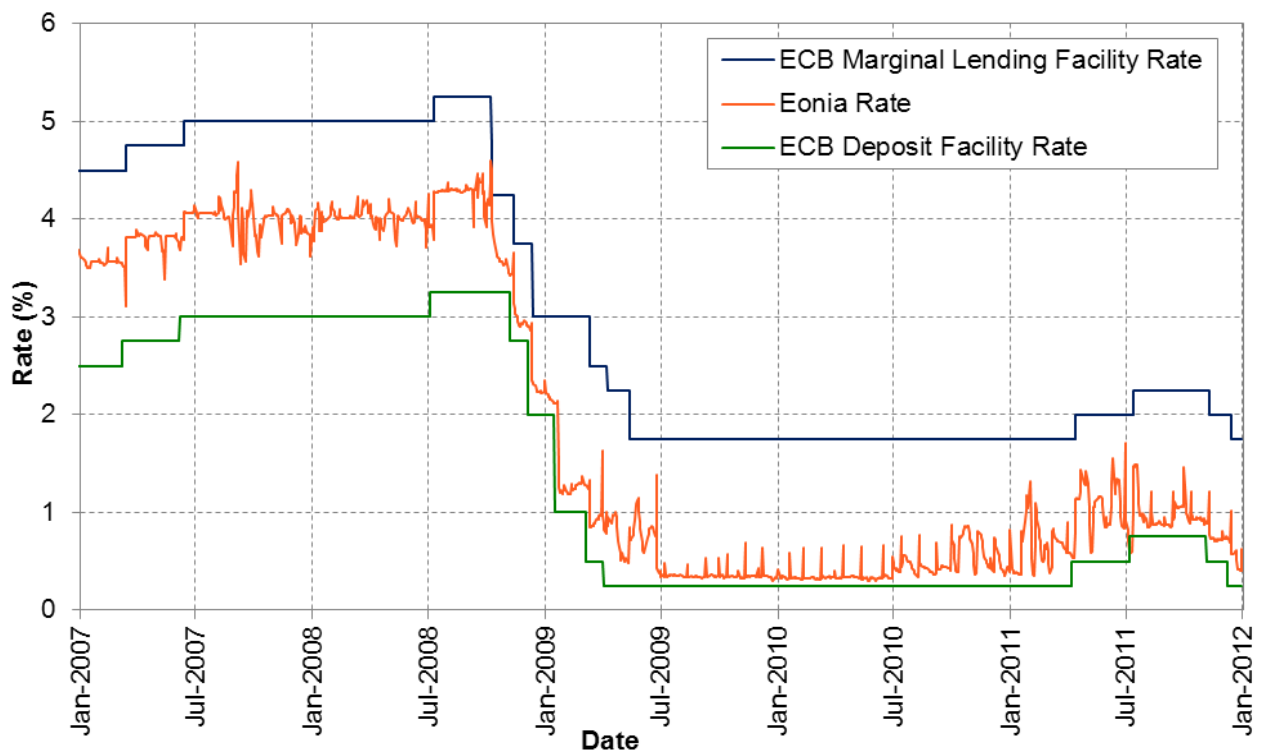

**Figure 2:** historical series of the Deposit Lending Facility rate, of the Marginal Lending Facility rate and of the Eonia rate (Jan. 2007 – Dec. 2011 window, sources: European Central Bank and Bloomberg).

Figure 1 reports the historical series of the Euribor Deposit 6 month (6M) rate versus the Eonia OIS 6 month (6M) rate over the time interval Jan. 2007 – Dec. 2011. During the crisis the solvency and solidity of the whole financial sector was brought into question and the credit and liquidity risk and premia associated to interbank transactions sharply increased. The Euribor rate dynamics immediately reflected these risk factors and raise to its highest levels over more than 10 years. As seen in Figure 1, the Euribor 6M rate suddenly increased on August 2007 and reached 5.49% on 10 October 2008, the maximum since its introduction in the 1999.

The historical trend of the Euribor – Eonia basis of Figure 1 can be divided in four distinct periods that are both related to different evolution phases of the financial turmoil.

The first covers the pre-crisis period that ends in August 2007, where both the credit and liquidity risk premia associated to interbank market participants were negligible and Euribor rates maintained levels close to the Eonia OIS ones.

The second covers the time interval from August 2007 to March 2009. During this period the interbank market was characterized by a simultaneous reduction of the OIS rate and an increase of the Euribor Deposit rate that lead to the explosion of the corresponding basis spread. The latter touches the peak of 222 bps in October 2008, when Lehman Brothers filed for bankruptcy protection and central banks decide to ease their policy cutting official rates.

The third period covers from March 2009 up to mid-2010 and it includes the phase of stabilization and reduction of the Euribor – Eonia basis, which maintained a level between 40 bps and 60 bps. After the failure of Lehman Brothers, central banks tried to fix markets' distress through the adoption of special policy measures that provided financial institutions with considerable amounts of liquidity, trying to ease the credit shortage and restore confidence within the interbank market.

The last period covers from the second half of 2010 to the end of 2011 and it is related to the sovereign crisis generated by some Euro zone state members. Financial markets were characterized by a strong sentiment of uncertainty related to the possible consequences that the failure of some European state could have in the Euro financial system. As we can observe from Figure 1, during this period the dynamics of the Euribor – Eonia basis is mainly driven by the decrease of the Eonia rate. Indeed, the Eonia OIS 6M market quote has experienced a sudden decrease of almost 90 bps between August and December 2011, while the Euribor Deposit 6M has displayed a reduction of just 20 bps during the same period. The decrease of the Eonia OIS rates was mainly a consequence of the intervention of the ECB that injected liquidity in the market allowing banks to fund themselves at lower rates than the ones of the interbank market.

The peculiar dynamics of the Euribor – Eonia basis can be ascribed to credit and liquidity risk factors reflected on unsecured money market rates. The increase of August 2007 experienced by the Euribor rates can be explained with an higher liquidity and credit risk premium required by the market over lending transactions with European interbank market participants. In section 2.4 we report some market evidences regarding the influence of the credit and liquidity risk factors within the new market's framework. In particular, we try to connect the explosion and movements of the Euribor – Eonia basis with two market proxies that, in our opinion, could help us to identify periods of credit and liquidity stress within the European interbank market.

Regarding the monetary policy effects, the intervention of central banks during the turmoil was finalized to restablishing and preserving an appropriate liquidity level in the interbank market. The most effective and common monetary policy instruments are referred to the "interest rate channel" set by central banks and tend to affect the short term money market rates like the Eonia rate, whose fixing is strictly connected with the two main ECB's standing facilities rates:

- The Deposit Facility rate: it is the official interest rate that the ECB offers to all the market eligible counterparties over overnight deposits. The Deposit Facility constitutes a liquidity absorption monetary policy instrument.

- The Marginal Lending Facility rate: it is the official interest rate that the ECB applies over overnight lending transaction with all the market eligible counterparties. The Marginal Lending Facility constitutes a liquidity providing monetary policy instrument.

The two standing facilities have the objective of steering the level of interbank overnight and short term rates and they defined the so-called "Rates Corridor". The Marginal Lending Facility rate is normally substantially higher than the corresponding money market rate and the Deposit Facility rate is usually substantially lower than the money market rate. Thus, financial institutions recur to the standing facilities in absence of any others convenient alternatives within the interbank market. Since there is no limit to the access of these standing facility, the Deposit Facility rate and the Marginal Facility rate define the overnight interest rate corridor that set a ceiling and a floor for the value of the Eonia rate. This is clear from Figure 2, showing that, over the period Jan. 2007 – Dec. 2011, Eonia is always higher than the Deposit Facility rate and lower than the Marginal Lending Facility rate. We notice that, since 2009, the Eonia rate is closer to the ECB Deposit Facility rate. This point will be explained in section 2.4, in connection with Figure 12.

Besides the conventional monetary policy instruments, the ECB introduced temporary monetary facilities such as fixed-rate refinancing operations with full-allotment, extension of the securities accepted as collateral and Long Term Refinancing Operations (LTROs) in order to ease the liquidity access among financial institutions. Monetary policy decisions affect also long term interest rates since they reflect expectations of the future evolution of short term interest rates. However, the impact of monetary policy decisions is less direct than those experienced by the Eonia rate and should be considered in terms of future growth expectations.

The Euribor – Eonia basis explosion plotted in Figure 1 is essentially a consequence of the different credit and liquidity risk reflected by Euribor and Eonia rates. We stress that such divergence is not a consequence of the counterparty risk carried by the financial contracts, Deposits and OISs, exchanged in the interbank market by risky counterparties, but depends on the different fixing levels of the underlying Euribor and Eonia rates. Clearly the market has learnt the lesson of the crisis and has not forgotten that these interest rates are driven by different credit and liquidity dynamics. From an historical point of view, we can compare this effect to the appearance of the volatility smile on the option markets after the 1987 crash (see e.g. Derman and Kani 1994). It is still there.

## 2.2. FRA Rates versus Forward Rates

The above considerations, referred to spot rates, related to Deposit and OIS contracts, apply to forward rates as well, related to Forward Rate Agreement (FRA) contracts. In Figure 4 we show the historical series of quoted Euribor FRA 6Mx12M rates versus the quoted Eonia FRA 6Mx12M rates, versus the Euribor forward rate 6Mx12M implied by the two quoted Deposits on Euribor 6M and Euribor 12M.

The Euribor FRA 6Mx12M rate is the equilibrium (fair) rate of a FRA contract starting at spot date (today + 2 working days in the Euro market), maturing in 12 months, with a floating leg indexed to the Euribor 6M rate, versus a fixed interest rate leg. At maturity, the floating leg pays the interest accrued with the Euribor 6M rate fixed 6 months before, over the time interval [6M, 12M]. The fixed leg pays the interest accrued with the fixed rate, over the same time interval. Thus the FRA equilibrium rate reflects the market expectations over the future fixing of the underlying Euribor 6Mx12M rate.

The Eonia FRA is similar to the Euribor FRA, but the floating leg is indexed to Eonia, daily fixed and compounded over the time interval [6M, 12M].

The Euribor forward rate $F(t, T_{i-1}, T_i)$, referred to the generic time interval $[T_{i-1}, T_i]$, is obtained through the standard formula (see e.g. Hull 2008),

$$F(t, T_{i-1}, T_i) := F_i(t) = \left(\frac{P(t, T_{i-1})}{P(t, T_i)} - 1\right) \frac{1}{\tau(T_{i-1}, T_i)}, \qquad (1)$$

where $P(t, T_i)$ is the price in $t$ of a zero-coupon bond maturing in $T_i$ and $\tau(T_{i-1}, T_i)$ represents the year fraction between $T_{i-1}$ and $T_i$. Equation 1 implicitly assumes that discounting from $T_i$ to $t$ at the spot rate $R(t, T_i)$ is equivalent to discounting from $T_i$ to $T_{i-1}$ at the corresponding forward rate $F(t, T_{i-1}, T_i)$ and then discounting from $T_{i-1}$ to $t$ at the spot rate $R(t, T_i)$.

Using equation 1, we compute the Euribor and Eonia forward rates as follows

$$\frac{1}{1+R_x^{Depo}(t,T_{i-1})\tau(t,T_{i-1})}\frac{1}{1+F_x^{FRA}(t,T_{i-1},T_i)\tau(T_{i-1},T_i)} = \frac{1}{1+R_y^{Depo}(t,T_i)\tau(t,T_i)}, \quad (2)$$

$$\frac{1}{1+R_d^{OIS}(t,T_{i-1})\tau(t,T_{i-1})}\frac{1}{1+F_d^{FRA}(t,T_{i-1},T_i)\tau(T_{i-1},T_i)} = \frac{1}{1+R_d^{OIS}(t,T_i)\tau(t,T_i)}, \quad (3)$$

where $R_x^{Depo}(t, T_i)$ is the market rate quoted at time $t$ for an Euribor Deposit with maturity $T_i$ with tenor $x$, $F_x^{FRA}(t, T_{i-1}, T_i)$ is the market rate quoted at time $t$ for an Euribor FRA contract covering the period $[T_{i-1}, T_i]$ with tenor $x$, $R_d^{OIS}(t, T_i)$ is the Eonia OIS market rate quoted at time $t$ with maturity $T_i$, $F_d^{FRA}(t, T_{i-1}, T_i)$ is the market rate quoted at time $t$ for a Eonia FRA contract covering the interval $[T_{i-1}, T_i]$ and the subscript $d$ refers to the discount curve. In Figure 3 we depict the mechanism associated with the two equations 2 and 3 above.

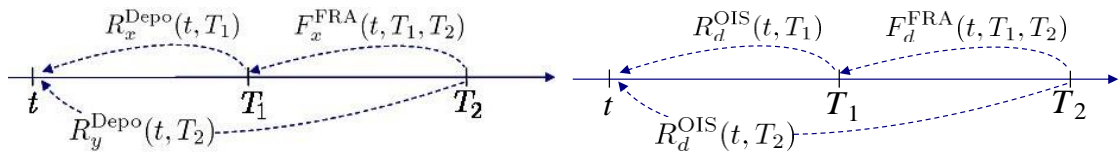

**Figure 3:** the two figures describe the mechanism we implemented in order to replicate the FRA market quotes.

In Table 1 we report a snapshot of the numbers obtained on 30 December 2011

Looking at the historical evolution in Figure 4, we observe that the three rates were essentially the same rate before the crisis, and diverged in August 2007, when the Forward, the Euribor FRA and the Eonia FRA acquired a positive basis with each other. The basis reached its maximum in October 2008, in correspondence of the Lehman Brothers' bankruptcy.

That difference can be justified by considering the nature of these rates. In particular, the credit an liquidity risk factors of FRA market equilibrium rates are mitigated by collateralization agreements that characterize FRA quoted contracts, while, in contrast, the deposit rates considered in the replication approach (i.e. Euribor Deposit 6M and Euribor Deposit 12M) are referred to unsecured transactions with different tenors (i.e. 6M and 12M) that, after the start of the crisis, reflect different liquidity and credit risk premia according to their maturity (Morini 2009).

Regarding the Eonia interest rate market, we compute the historical series of the Eonia Forward 6Mx12M rate during the interval Jan. 2007 – Dec. 2011 according to the equation 3 and we reported the results in Figure 5. We notice that the difference between the Eonia FRA 6Mx12M market rates and the corresponding forward rate is negligible over the whole observation period (average difference of 0.7 bps in absolute terms).

The Eonia OIS rates used for the FRA replica are obtained through the compounding of the Eonia O/N rate. Hence, the credit and liquidity risk components carried by the Eonia Forward rates can be considered negligible and consistent with the risk premia reflected by the Eonia FRA market rates.

In section 2.5 we report some findings of Mercurio (2009) who has proven that the above effects may be explained within a simple credit model that considers a default-free zero coupon bond and a risky zero coupon bond emitted by a defaultable counterparty.

**Eonia FRA Replication (30 Dec. 2011)**

| Eonia OIS Maturity | Eonia OIS Quote (%) | Eonia FRA Start/End Dates | Eonia FRA Quote (%) | Eonia FRA Replica (%) | Difference Replica-Quote (bps) |
|---|---|---|---|---|---|
| 1M | 0.396 | 1Mx2M | 0.392 | 0.392 | 0.0 |
| 2M | 0.394 | 2Mx3M | 0.386 | 0.385 | -0.1 |
| 3M | 0.391 | 1Mx4M | 0.383 | 0.382 | -0.1 |
| 4M | 0.386 | 2Mx5M | 0.371 | 0.370 | -0.1 |
| 5M | 0.380 | 3Mx6M | 0.370 | 0.371 | 0.1 |
| 6M | 0.381 | 6Mx12M | 0.372 | 0.372 | 0.0 |
| 12M | 0.376 | | | | |

**Euribor FRA Replication (30 Dec. 2011)**

| Euribor Depo. Maturity | Euribor Deposit Quote (%) | Euribor FRA | Euribor FRA Quote (mid, %) | Euribor FRA Replica (%) | Difference Replica-Quote (bps) |
|---|---|---|---|---|---|
| 1M | 0.980 | 1Mx4M | 1.223 | 1.500 | 27.7 |
| 2M | 1.150 | 2Mx5M | 1.130 | 1.677 | 54.7 |
| 3M | 1.310 | 3Mx6M | 1.067 | 1.804 | 73.7 |
| 4M | 1.380 | 4Mx7M | 1.016 | 1.948 | 93.2 |
| 5M | 1.460 | 5Mx8M | 0.964 | 2.080 | 111.6 |
| 6M | 1.560 | 6Mx9M | 0.931 | 2.103 | 117.2 |
| 7M | 1.620 | 1Mx7M | 1.471 | 1.728 | 25.7 |
| 8M | 1.690 | 2Mx8M | 1.365 | 1.883 | 51.8 |
| 9M | 1.740 | 3Mx9M | 1.292 | 1.958 | 66.6 |
| 10M | 1.790 | 4Mx10M | 1.246 | 2.073 | 82.7 |
| 11M | 1.840 | 5Mx11M | 1.200 | 2.154 | 95.4 |
| 12M | 1.900 | 6Mx12M | 1.172 | 2.243 | 107.1 |
| 18M | 1.860 | 12Mx18M | 1.125 | 1.736 | 61.1 |
| 24M | 1.870 | 18Mx24M | 1.224 | 1.868 | 64.4 |
| | | 12Mx24M | 1.481 | 1.800 | 31.9 |

**Table 1**: top panel: comparison between Eonia OIS, FRA and Forward rates for several start/end dates quoted in the market. Bottom panel: the same for Euribor Deposits, FRA and Forward rates. Note that Euribor FRA contracts have different underlying rate tenors: 1Mx4M – 6Mx9M are indexed to the Euribor 3M, 1Mx7M – 18Mx24M are indexed to Euribor 6M, and 12Mx24M is indexed to the Euribor 12M (source: ICAP, reference date: 30 Dec. 2011).

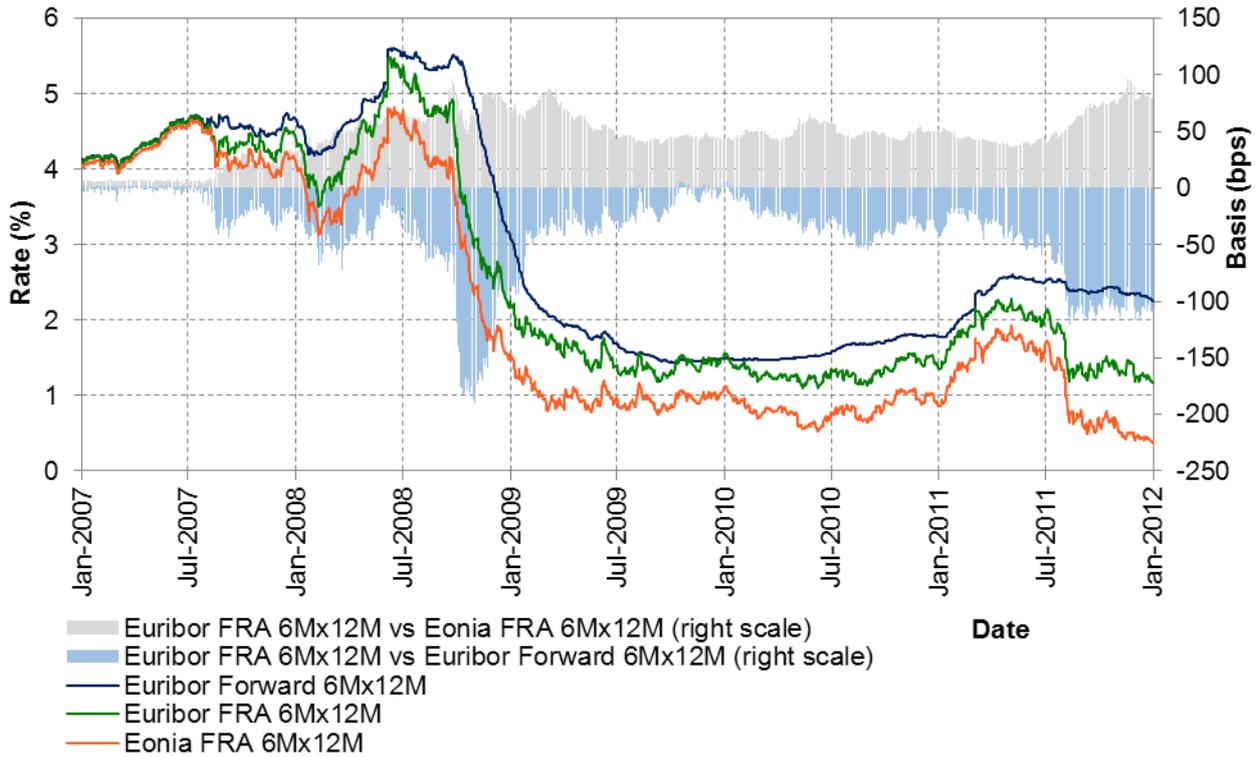

**Figure 4**: Euribor FRA 6Mx12M market rate versus Eonia FRA 6Mx12M market rate versus Euribor Forward 6Mx12M rate (computed using equation 2). The corresponding spreads are shown on the right y-axis (Jan. 2007 – Dec. 2011 window, source: Bloomberg).

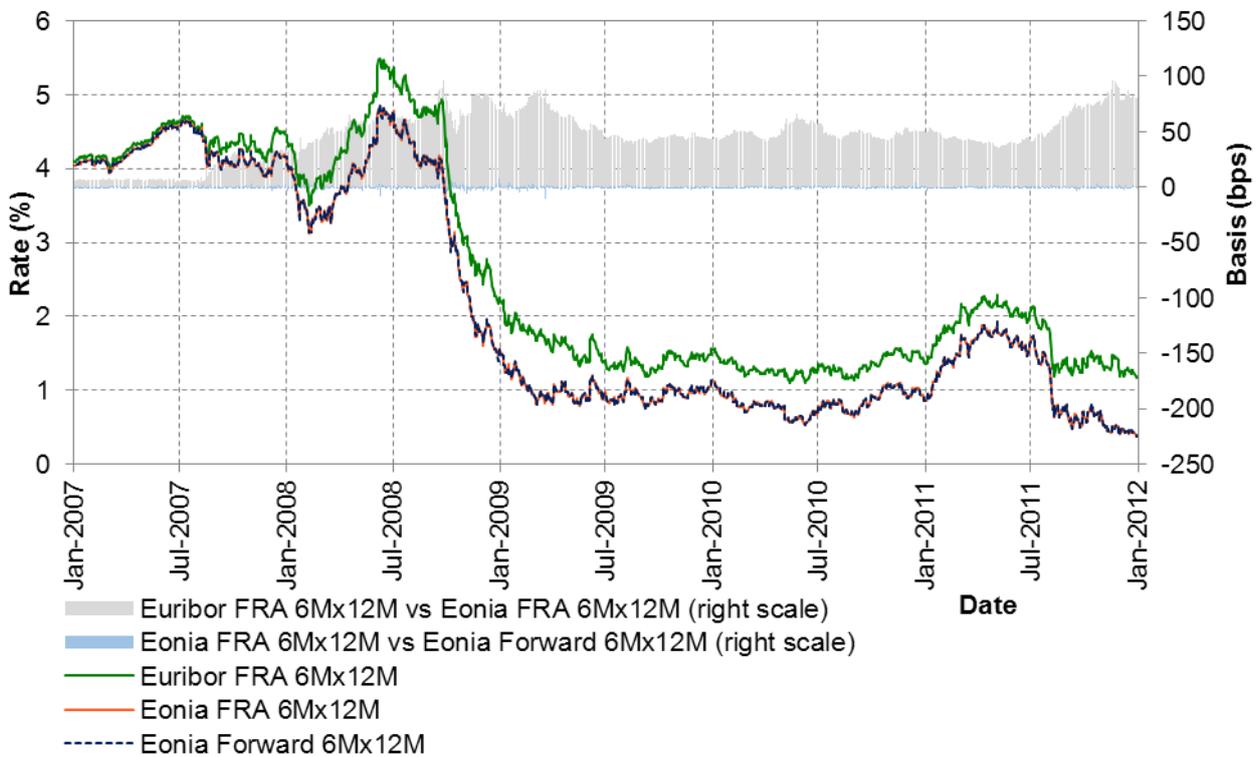

**Figure 5**: he Euribor FRA 6Mx12M market rate versus Eonia FRA 6Mx12M market rate versus Eonia Forward 6Mx12M rate (computed using equation 3). The corresponding spreads are shown on the right y-axis (Jan. 2007 – Dec. 2011 window, source: Bloomberg).

## 2.3. Basis Swaps

A third evidence of the regime change after the credit crunch is the explosion of the Basis Swaps spreads. In Figure 6 we report three historical series of quoted Basis Swap equilibrium spread, Euribor 3M vs Euribor 6M, Euribor 6M vs Euribor 12M, Euribor 3M vs Eonia, all at 5 years swap maturity. Basis Swaps are quoted on the Euro interbank market in terms of the difference between the fixed equilibrium swap rates of two swaps. For instance, the quoted Euribor 3M vs Euribor 6M Basis Swap rate is the difference between the equilibrium swap rates of a first standard swap with an Euribor 3M floating leg (quarterly frequency) vs a fixed leg (annual frequency), and of a second swap with an Euribor 6M floating leg (semi-annual frequency) vs a fixed leg (annual frequency). The frequency of the floating legs is the "tenor" of the corresponding Euribor rates. The Eonia floating legs are indexed to the shortest tenor rate (1 day), have annual frequency, and the floating coupon rate is given by the simple composition of the Eonia rates fixed daily during the coupon period.

As we can see in Figure 6, the Basis Swap spreads were negligible (or even not quoted) before the crisis. They suddenly diverged in August 2007 and peaked in October 2008 with the Lehman crash. Figure 7 reports spot Basis Swap spreads (reference date 30/12/2011) for different pairs of rates on several maturities. Basis Swap spreads not directly observable on the market have been computed from market quotations.

The Basis Swap involves a sequence of FRA rates carrying the credit and liquidity risk discussed in sections 2.1 and 2.2 above. Hence, the basis spread explosion can be interpreted in terms of the different credit and liquidity risk carried by the underlying FRA rates with different tenors, as in Figure 4. In Figure 6 and Figure 7 we see another example that, after the crisis, a swap floating leg indexed to the higher rate tenor (e.g. 6M) has an higher value with respect to the floating leg indexed to the shorter rate tenor (3M) with the same maturity, thus a positive spread emerges between the two corresponding equilibrium rates (or, in other words, a positive spread must be added to the 3M floating leg to equate the value of the 6M floating leg). In Figure 7 we observe that the magnitude of the Basis Swap spread increases with the tenor difference (see Bianchetti 2010 and Bianchetti 2011).

According to Morini (2009), a Basis Swap between two interbank counterparties under collateral agreement can be described as the difference between two investment strategies. Fixing, for instance, a Basis Swap Euribor 3M vs Euribor 6M with 6M maturity, scheduled on 3 dates $T_0$, $T_1 = T_0 + 3M$, $T_2 = T_0 + 6M$, we have the following two strategies:

1. 6M floating leg: at $T_0$ choose a counterparty $C_1$ with an high credit standing (that is, belonging to the Euribor Contribution Panel) with collateral agreement in place, and lend the notional for 6 months at the Euribor 6M rate prevailing at $T_0$ (Euribor 6M flat because $C_1$ is an Euribor counterparty). At maturity $T_2$ recover notional plus interest from $C_1$. Notice that if counterparty $C_1$ defaults within 6 months we gain full recovery thanks to the collateral agreement.

2. 3M+3M floating leg: at $T_0$ choose a counterparty $C_1$ with an high credit standing (belonging to the Euribor Contribution Panel) with collateral agreement in place, and lend the notional for 3 months at the Euribor 3M rate (flat) prevailing at $T_0$. At $T_1$ recover notional plus interest and check the credit standing of $C_1$: if $C_1$ has maintained its credit standing (it still belongs to the Euribor Contribution Panel), then lend the money again to $C_1$ for 3 months at the Euribor 3M rate (flat) prevailing at $T_1$, otherwise choose another counterparty $C_2$ belonging to the Euribor Panel with collateral agreement in place, and lend the money to $C_2$ at the same interest rate. At maturity $T_2$ recover notional plus interest from $C_1$ or $C_2$. Again, if counterparties $C_1$ or $C_2$ defaults within 6 months we gain full recovery thanks to the collateral agreements.

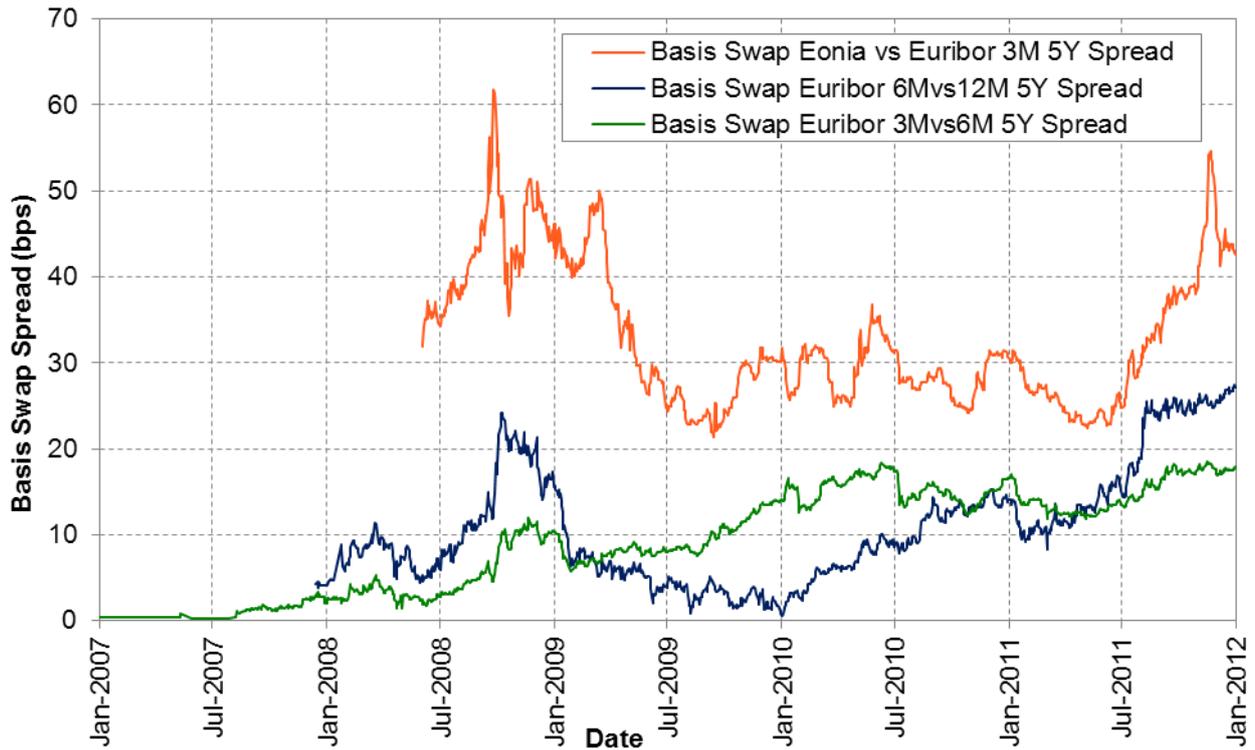

**Figure 6:** Basis Swap spreads: Euribor 3M Vs Euribor 6M, Euribor 6M Vs Euribor 12M and Eonia Vs Euribor 3M (Jan. 2007 – Dec. 2011 window, source: Bloomberg). All the quotations present a maturity of 5Y. Notice that the daily market quotations for some Basis Swap were not even available before the crisis.

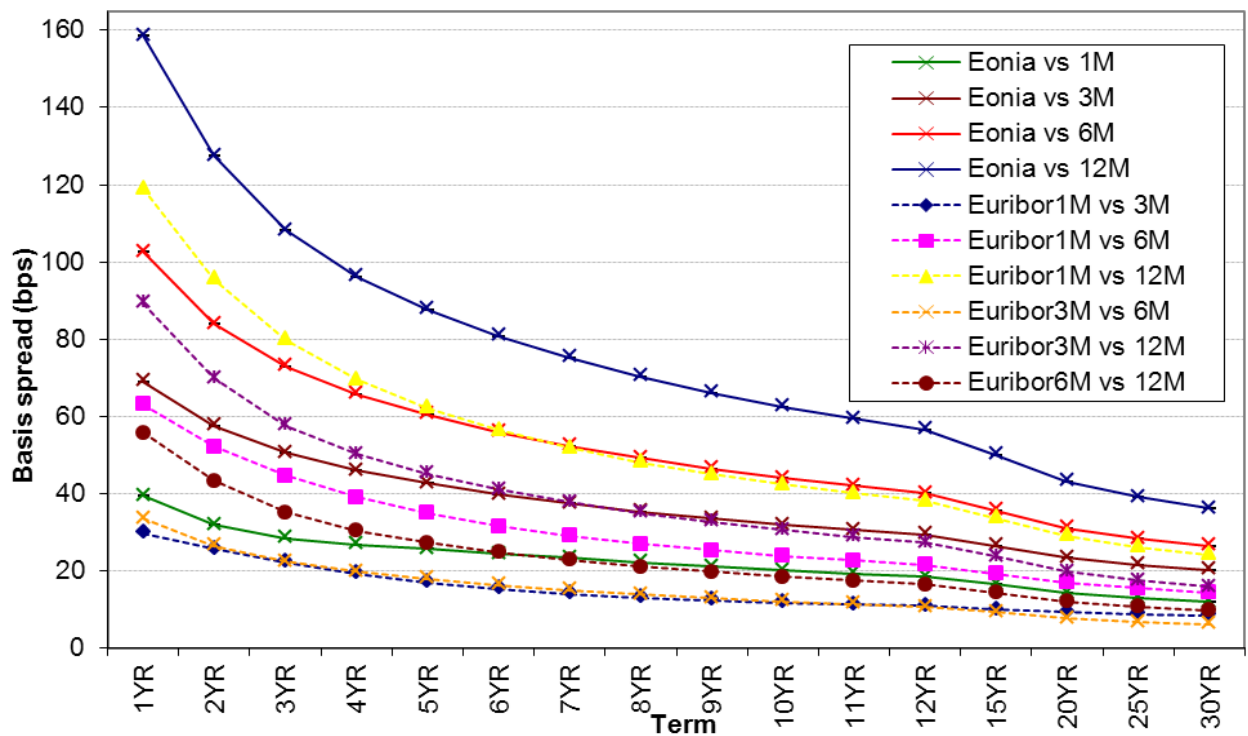

**Figure 7:** Basis Swap spreads Eonia vs Euribor $x$M and Euribor $x$M vs $y$M over several maturities (reference date: 30/12/2011, source: Reuters).

Clearly, the 3M+3M leg implicitly embeds a bias towards the group of banks with the best credit standing, typically those belonging to the Euribor Contribution Panel. Hence, the credit risk carried by the 3M+3M leg must be lower than that carried by the 6M leg. In other words, the expected survival probability of the borrower in the 3M leg in the second 3M-6M period is higher than the survival probability of the borrower in the 6M leg in the same period. This lower risk is embedded into lower Euribor 3M + 3M rates with respect to Euribor 6M rates. But with collateralization the two legs have both negligible counterparty risk. Thus a positive spread must be added to the 3M+3M leg to reach equilibrium. The same discussion can be repeated, mutatis mutandis, in terms of liquidity risk.

In Figure 8 we show a pictorial view of floating legs indexed to rates with different tenors. In equation 4 we report the corresponding leg values.

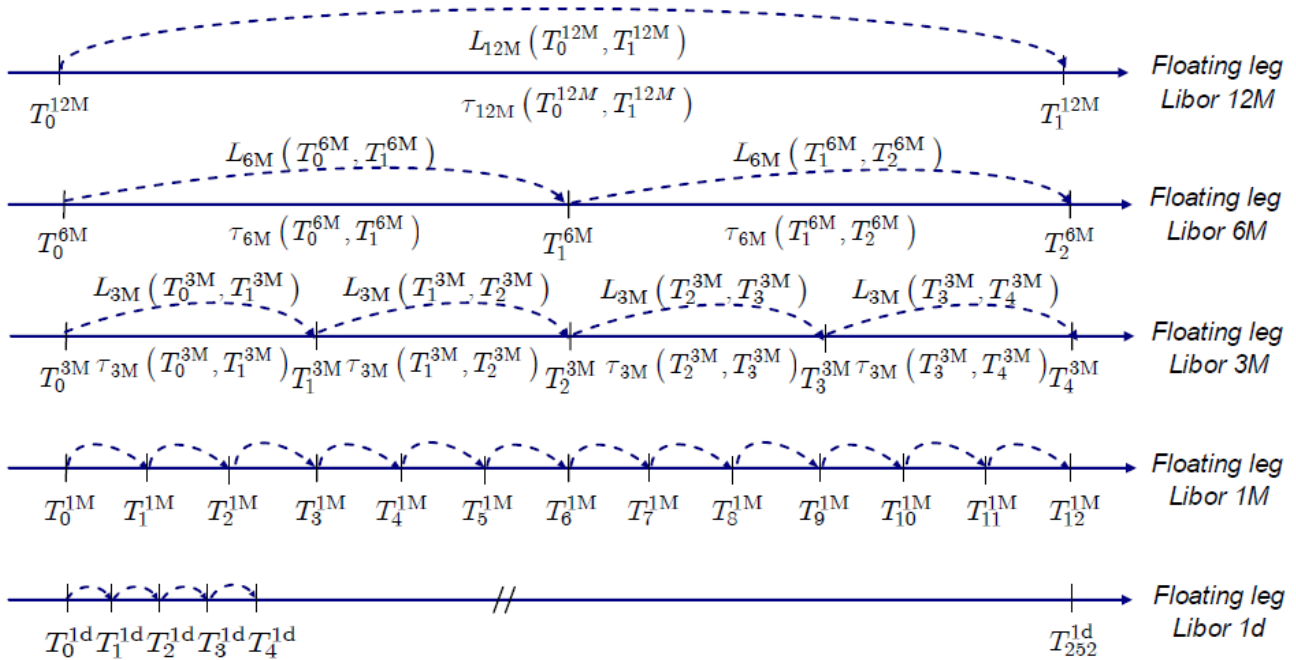

**Figure 8:** representation of floating Swap legs with different tenors (12M, 6M, 3M, 1M, 1d).

Before the financial crisis, since the liquidity and credit risk embedded in interbank rates with different tenors were very similar (and small), stream of cash flows with same maturity but different tenors could be replicated one with each others, and all these floating legs had the same value. The start of the financial turmoil and the consequent explosion of Basis Swap spread between rates with different tenors has invalidated classical no-arbitrage relations, such as equations 1 and 4, such that these floating legs acquired different values

The start of the financial turmoil and the consequently explosion of Basis Swap spread between rates with different tenors has invalidated classical no-arbitrage relations that, such as the following one, do not hold anymore,

$$\mathbf{Swap}_{12M}(t,\mathbf{T}) = P_d(t,T_{12M})L_{12M}(t,T_{12M})\tau_{12M}(t,T_{12M})$$

$$\neq \mathbf{Swap}_{6M}(t,\mathbf{T}) = \sum_{i}^{2} P_d(t,T_i)F_{6M,i}(t)\tau_{6M}(T_{i-1},T_i)$$

$$\neq \mathbf{Swap}_{3M}(t,\mathbf{T}) = \sum_{i}^{4} P_d(t,T_i)F_{3M,i}(t)\tau_{3M}(T_{i-1},T_i)$$

$$\neq \mathbf{Swap}_{1M}(t,\mathbf{T}) = \sum_{i}^{12} P_d(t,T_i)F_{1M,i}(t)\tau_{1M}(T_{i-1},T_i)$$

$$\neq \mathbf{Swap}_{1d}(t,\mathbf{T}) = \sum_{i}^{252} P_d(t,T_i)F_{1d,i}(t)\tau_{1d}(T_{i-1},T_i)$$

$$\neq 1 - P_d(t,T_{1Y})$$

$$\text{with}\quad F_{x,i}(t) = \left(\frac{P(t,T_{i-1})}{P(t,T_i)} - 1\right)\frac{1}{\tau(T_{i-1},T_i)} \quad (4)$$

where $\text{Swap}_x(t,\mathbf{T})$ represents the net present value of the floating leg of a Swap indexed to the Euribor rate $L_x(t,T_x)$ with tenor $x$ and with payment times according to the dates set $\mathbf{T} = \{T_0, \ldots, T_n\}$, $F_{x,i}(t)$ is the FRA market rate referred to the period $[T_{i-1}, T_i]$ associated to the Euribor rate with tenor $x$ (i.e. $x = 1d, 1M, 3M, 6M, 12M$).

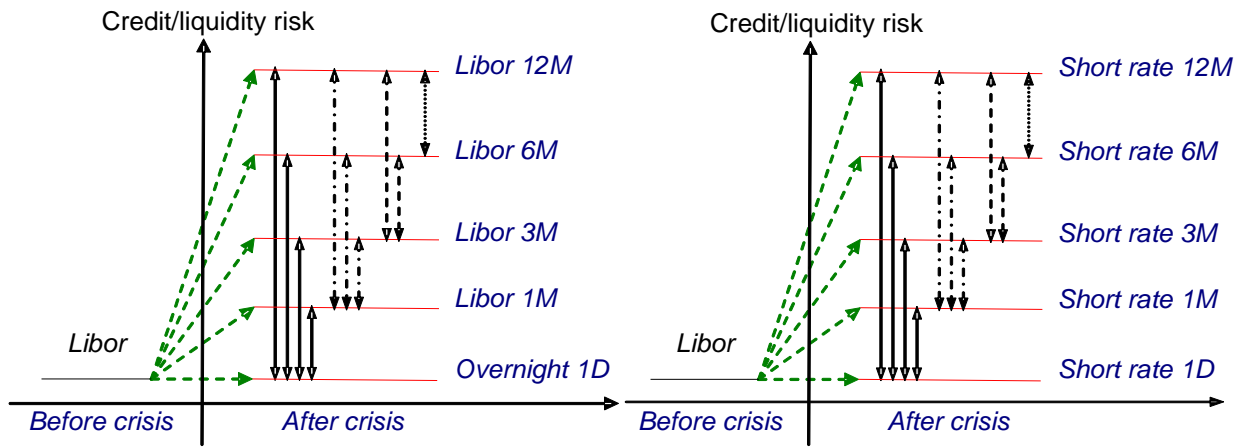

**Figure 9:** representation of the interest rate market segmentation.

We stress that the credit and liquidity risk involved here are those carried by the risky Libor rates underlying the Basis Swap, reflecting the average default and liquidity risk of the interbank money market (of the Libor panel banks), not those associated to the specific counterparties involved in the financial contract. We point out also that such effects were already present before the credit crunch, as discussed e.g. in Tuckman and Porfirio (2004), and well known to market players, but not effective due to negligible basis spreads.

## 2.4. The Credit and Liquidity Risk Components

In this section we try to highlight the credit and liquidity risk impact on the historical trend of the Euribor – Eonia basis during the period Jan. 2007 – Dec. 2011. To this aim, we introduce two different indexes that help us to underline credit and liquidity market stress periods.

Regarding the credit risk, we build an index representative of the credit risk in the European financial sector that we call **Synthetic CDS Euribor Index**. This index considers daily quotations of 5 years maturity CDS spread referred to financial institutions that belong to the Euribor panel in December 2011. Its computation replicates the fixing mechanism of the Euribor rates. Hence, for each reference date, we exclude the highest and the lowest 15% CDS spread quotations and compute the average of the remaining 70% quotes. The Synthetic CDS Euribor Index thus represents the average cost for protection against the default of a Libor panel bank within the European financial market.

Regarding the liquidity risk, we compute an index, called **Liquidity Surplus Index**, that considers official data reported by the ECB. The index is given by the sum of the total amount of the deposits posted by the EU financial institutions at the ECB's Deposit Facility and of the current account holdings exceeding the EUR market-wide level minimum reserve requirement that are held by EU financial institutions at the ECB. We refer to this aggregation as a proxy of the liquidity surplus in the Euro zone interbank market.

The ECB requires credit and financial institutions to hold minimum reserves amounts on accounts managed by National Central Banks. The minimum reserve system has the scope of stabilizing the market interest rates and to facilitate the role of the ECB as liquidity supplier for the interbank market. The amount of minimum reserves is fixed, on a monthly basis, according to each financial institution's reserve base and the compliance of the requirements is verified considering the average, during a certain maintenance period, of the amounts posted at the reserve accounts. This mechanism ensures flexibility to financial institutions that can face minimum reserves provisions without compromising their business or investing opportunities. Holdings of required reserves are remunerated at the Main Refinancing Operation (MRO) rate, while holdings that exceeded the reserve requirement are free of remuneration. The minimum reserve is a liquidity absorption standing facility (ECB 2010).

The amounts posted by financial institutions at the Deposit Facility and the Excess Reserves help us to track liquidity stress of financial markets. Indeed, the higher is the Liquidity Surplus Index, the stronger is the preference to deposit cash reserves at the ECB instead of lending in the interbank market or investing in more profitable (and risky) activities.

In Figure 10 we report the historical series of the Synthetic CDS Euribor Index vs the Euribor 6M – Eonia OIS 6M basis, of the market quotes of the Basis Swap Eonia Vs Euribor 3M and of the Synthetic CDS Euribor Index over the period Jan. 2007 – Dec. 2011. We can observe that the Synthetic CDS Euribor Index reached a first peak in August 2007 in relation to the rise of concerns over banks' exposure to credit structured products (i.e. CDO, ABS etc.). This first increase of the premia against the default of primary financial institutions matched the explosion of the Euribor – Eonia basis (Figure 10 – spot 1) and it highlights a generalized growth of the default risk perceived in the interbank market reflected by an increase of the Euribor rates (see Figure 1).

Since then. the index started to increase rapidly and maintaining an upward trend over the whole time interval we considered. The second and third peak of the Synthetic CDS Euribor Index are related to the bail-out of the investment bank Bear Stearns (14 March 2008, Figure 10 – spot 2) and to the bankruptcy of Lehman Brothers (15 September 2008, Figure 10 – spot 3) respectively. The market uncertainty related to these two periods corresponds to an increase of the Euribor – Eonia basis. Before that a period of market relax occurred in 2009, the Synthetic CDS Euribor Index reached a fourth peak (March 2009, Figure 10 – spot 4) due to the deterioration of financial markets unlashed by the failure of Lehman Brothers. This increase of the credit risk perceived by the market is not reflected by a similar increase in terms of magnitude of the Euribor – Eonia basis that was mainly driven by the loosening monetary policy decisions of central banks. The Synthetic CDS Euribor Index reached its maximum during September 2011 (Figure 10 – spot 5), in correspondence of Italy's credit rating cut, and it was very far from the pre-crisis level when banks where considered "too big to fail". This rise in the credit risk was matched by an increase of the Euribor – Eonia basis that reached 127 bps on 01 December 2011 (Figure 10 – spot 6).

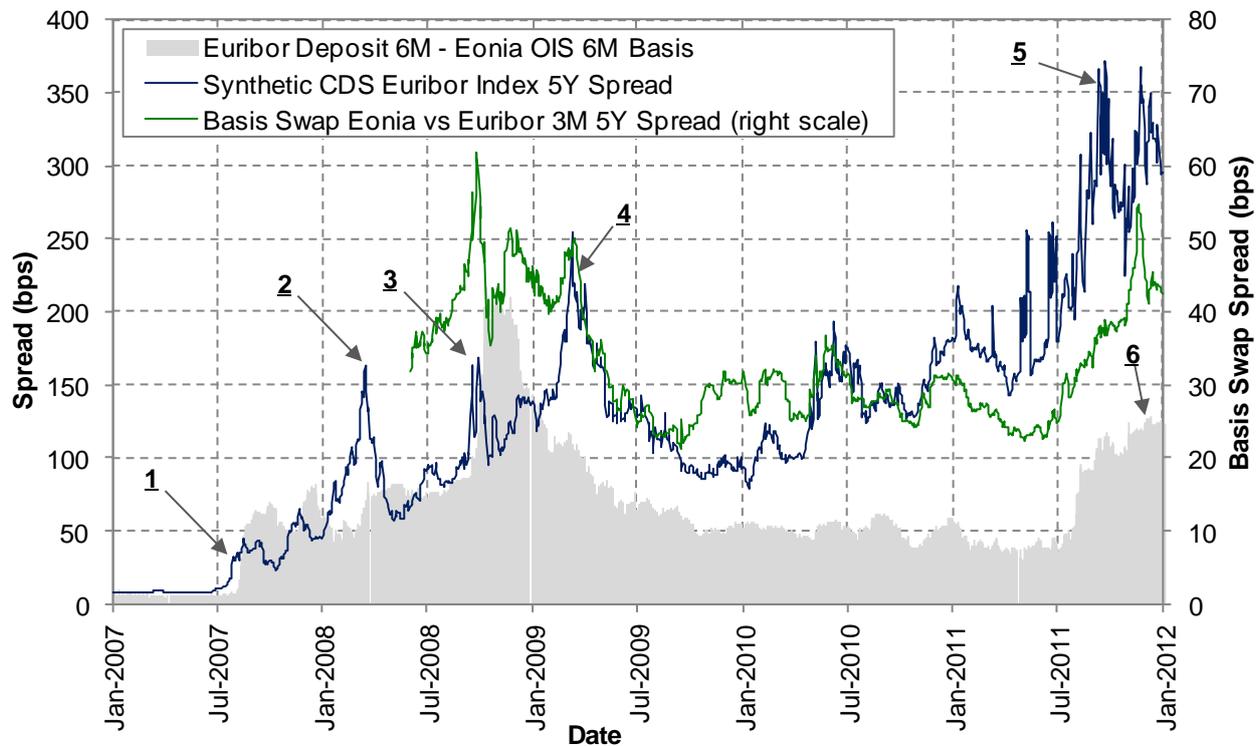

**Figure 10:** Synthetic CDS Euribor Index 5Y spread (line), Basis Swap Eonia Vs Euribor 3M 5Y spread (line, right scale) and Euribor Deposit 6M – Eonia OIS 6M basis (area) from Figure 1 (Jan. 2007 – Dec. 2011 window, sources: Bloomberg and ICAP). Notice that the basis swap has the same payment frequency (3 months) and maturity (5Y) of the Synthetic CDS Index.

Regarding the relation between the Basis Swap Eonia Vs Euribor 3M and the Synthetic CDS Euribor Index, we can observe a close trend of the two historical series, especially in correspondence of an increase of the credit risk perceived in the interbank market. The increase of the Basis Swap spread in correspondence of a rise in the average default risk seems to reveal a stronger relevance of the credit risk component over longer maturities (i.e. 5Y).

The liquidity risk component in Euribor and Eonia interbank rates is distinct but strongly related to the credit risk component. According to Acerbi and Scandolo (2007), liquidity risk may appear in at least three circumstances:

1. lack of liquidity to cover short term debt obligations (funding liquidity risk),
2. difficulty to liquidate assets on the market due excessive bid-offer spreads (market liquidity risk),
3. difficulty to borrow funds on the market due to excessive funding cost (systemic liquidity risk).

Following Morini (2009), these three elements are, in principle, not a problem until they do not appear together, because a bank with, for instance, problem 1 and 2 (or 3) will be able to finance itself by borrowing funds (or liquidating assets) on the market. During the crisis these three scenarios manifested themselves jointly at the same time, thus generating a systemic lack of liquidity (see e.g. Michaud and Upper 2008).

Clearly, it is difficult to disentangle liquidity and credit risk components in the Euribor and Eonia rates, because, in particular, they do not refer to the default risk of one counterparty in a single derivative deal but to a money market with bilateral credit risk (see the discussion in Morini (2009) and references therein).

In the Euro system the ECB is responsible of ensuring and maintaining the liquidity of the financial market through several monetary facilities and open market operations. Any liquidity injection in the interbank market should be absorbed by financial institutions. Before the financial turmoil, the liquidity provided by the ECB aimed mainly to satisfy the market's liquidity needs and banks could rely on an easy and convenient access to the interbank market for their short term liquidity operations.

In Figure 11 we compare the historical trend of the Synthetic CDS Euribor Index, of the Liquidity Surplus Index, of the Euribor 6M – Eonia OIS 6M basis and of the Basis Swap Eonia Vs Euribor 3M. The first main intervention of the ECB during the financial crisis was in October 2008 (Figure 11 – spot 1) an it regarded the adoption of several measures such as the cut of the official interests rates in conjunction with others central banks[3], the introduction of a fixed-rate refinancing operation with full-allotment, the extension of the securities accepted as collateral by the central bank and the increase of the number of financial institutions that can accede to the ECB monetary policy channels (ECB 2008a, 2008b, 2008c, 2008d). As we can observe in Figure 11, the new monetary policy decisions put in force by the ECB led to a sudden explosion of the Liquidity Surplus that exactly matches the most relevant increase experienced by the Euribor – Eonia spread during the period Jan. 2007 – Dec. 2011.

The Liquidity Surplus Index reached a second peak on the 25 June 2009 (Figure 11 – spot 2) and its increase is due to the introduction by the central bank of a LTRO with a 12M term (ECB 2009a). This non-standard facility provided the European financial market with an unlimited amount of liquidity with 1 year maturity. This intervention reduced the liquidity shortage of the market and it is accompanied by a reduction of the Euribor – Eonia basis and a decrease of the credit risk reflected by the Synthetic CDS Euribor Index.

At the end of the 2009 (Figure 11 – spot 3) we can notice a third jump in the liquidity amount posted at the ECB. This sudden variation can be ascribed to the extension of the fixed-rate refinancing operations with full allotment introduced in October 2008 (ECB 2009b).

From Figure 11 we note that the Liquidity Surplus index experienced an upward trend during the period Jan. 2010 – Jul. 2010 that is exacerbated in May 2010 by the worsening of the so called "sovereign debt crisis" related to market concerns over Greece's capability to maintain its debt obligations. The effects of this market uncertainty are reflected also by the Synthetic CDS Euribor index that increased almost up to 200 bps in June 2010. The LTRO introduced in June 2009 by the ECB expired at the end of May 2010 (Figure 11 – spot 4). The amount of liquidity surplus shrank significantly and the Liquidity Surplus Index decreased until July 2011.

During the period Jul. 2009 – Jul 2011 the Euribor – Eonia basis has maintained a relative stable and low level compared to Aug. 2007 – Jun. 2009, showing contained peaks in correspondence of a simultaneous increase of both the Liquidity Surplus Index and the Synthetic Surplus Index.

The second half of the 2011 was characterized by the second phase of the sovereign debt crisis that started to affect countries such as Italy and Spain. The market conditions in terms of credit and liquidity risk deteriorated significantly and that was promptly reflected by the increase of both the Liquidity Surplus Index and the Synthetic CDS Euribor Index. In this period the Eonia OIS rates decreased significantly and Euribor rates remained almost stable (see Figure 1) leading to a relevant rise of the correspondent basis.

The Liquidity Surplus Index reached its maximum on the 22$^{nd}$ December 2011 (Figure 11 – spot 5) in correspondence of the ECB's decision to put in place a multi-tranche LTRO with a maturity of 3 years (ECB 2011). The first LTRO tranche, which took place in the 21 December 2011, provided €489.2 billion to 523 financial institutions (ECB 2012) The next day the Liquidity Surplus Index hit a value of €483 billion and almost all the liquidity offered to the market by the ECB was posted at the ECB's accounts. The difficult credit and liquidity conditions experienced by the European financial market in the last half of 2011 led to a significant increase in the Euribor – Eonia basis.

---

[3] The Bank of Canada, the Bank of England, the Federal Reserve, Sveriges Riksbank and the Swiss National Bank.

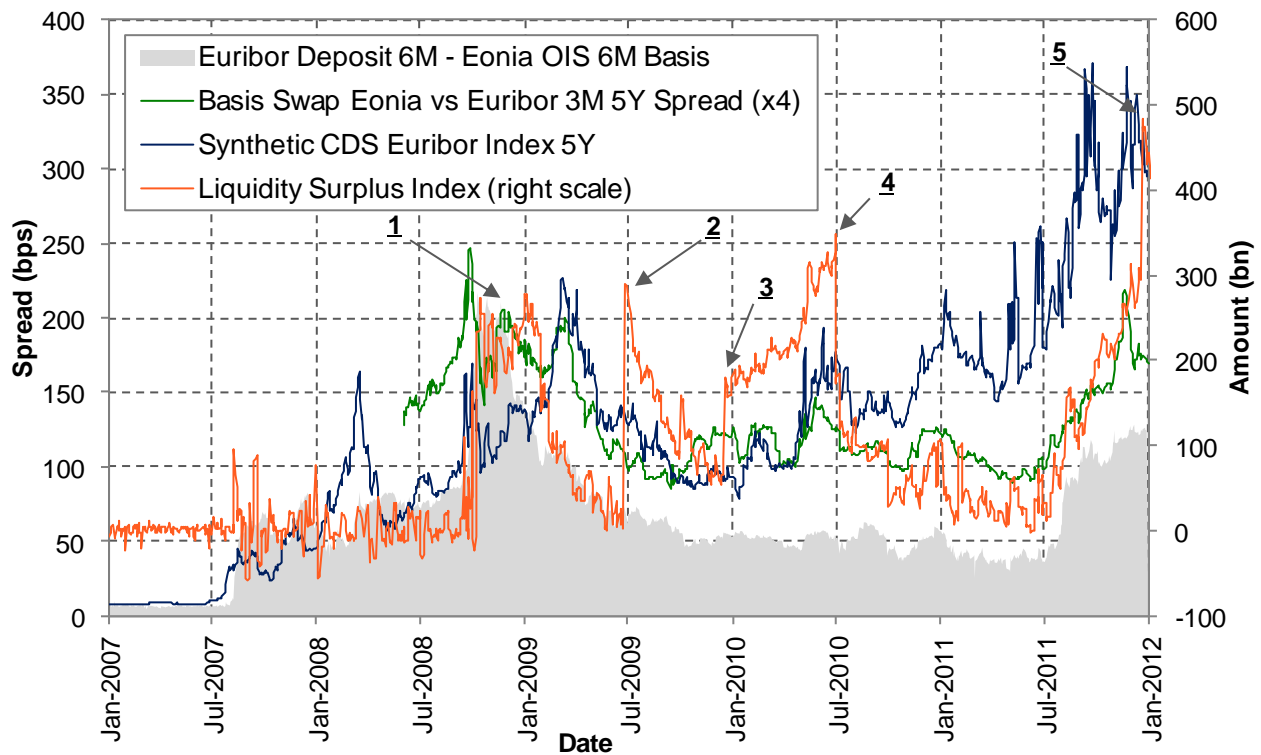

**Figure 11:** Synthetic CDS Euribor index 5Y (line), Liquidity Surplus Index of the European interbank market (line, right scale), Basis Swap Eonia Vs Euribor 3M 5Y (line) and Euribor Deposit 6M – Eonia OIS 6M basis (area). We multiplied by 4 the Basis Swap Eonia Vs Euribor 3M 5Y to better compare its historical trend (window Jan. 2007 – Dec. 2011, sources: Bloomberg and ECB).

In Figure 12 we compare the trend of the Liquidity Surplus index and of the Eonia Volume. We can observe that an increase in the liquidity amount posted at the ECB is always accompanied by a reduction of the total amount traded in the European money market. The drain of liquidity that affected the money market is the main reason of the closeness of the Eonia rate to the Deposit Facility rate during the financial crisis (see Figure 2).

By considering the trend of the Liquidity Surplus Index we argue that from the Lehman Brothers' bankruptcy up to the end of the 2011 the liquidity risk factor has played a key role, in conjunction with the credit risk, in explaining the trend of the Euribor – Eonia basis. Generally, we can observe an increase of the difference between Euribor and Eonia OIS rates when both the Synthetic CDS Euribor Index and the Liquidity surplus Index start to go up. This combined upward movement reveals an increase of the overall risk perceived within the interbank market. Observing the historical series reported in Figure 11, we claim that the Euribor – Eonia basis' peak of October 2008 is caused, initially, by the increase in the average default risk of the market in correspondence of the Lehman crash and, subsequently, by the liquidity risk in the interbank market and the drastic official interest rate cut operated by the ECB during that period. Also in the second half of 2011, the upward trend of the Euribor – Eonia basis was driven by a simultaneous rise of both the credit and liquidity risk in the interbank market, reflected in the market through a decrease of the Eonia OIS rates and almost stable Euribor rates.

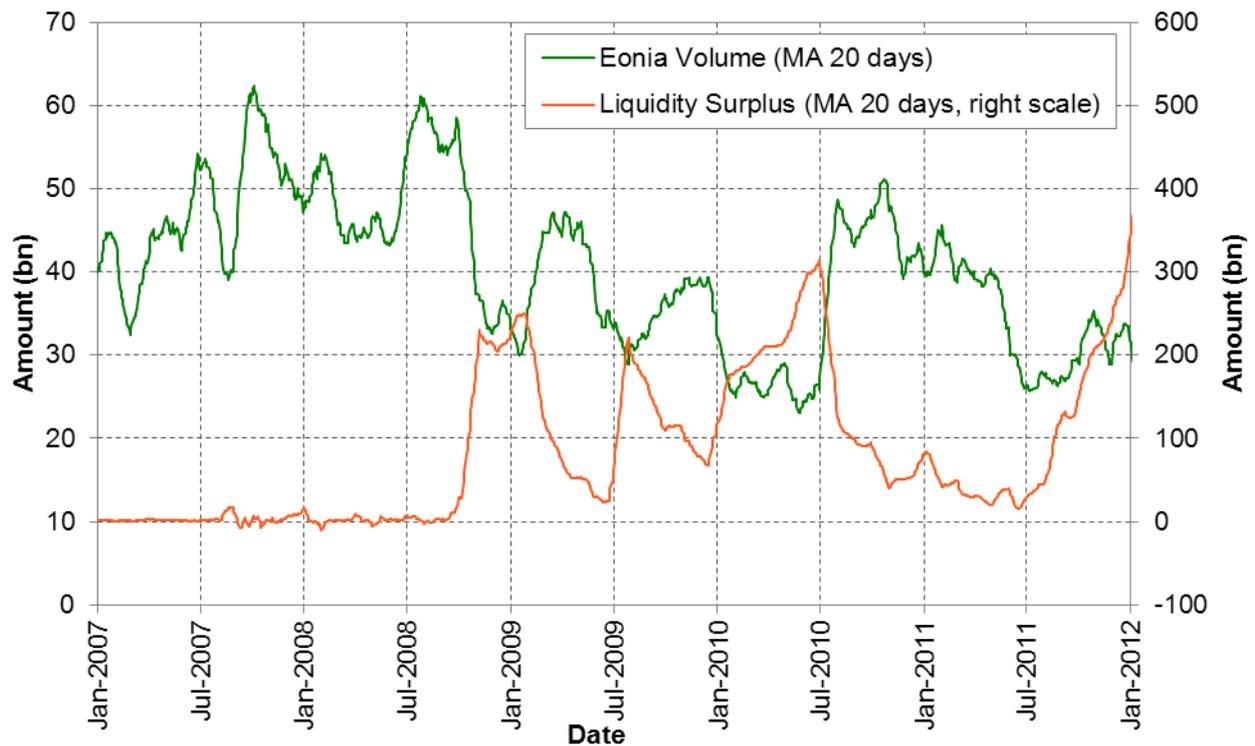

**Figure 12:** moving average 20 days of the Liquidity Surplus index and of the Eonia Volume (window Jan. 2007 – Dec. 2011). Note that we reported the Eonia Volume on the right scale since the two historical series present different magnitudes. (Sources: Bloomberg, ECB).

## 2.5. A Simple Credit Model

In order to explain the basis divergence after the credit crunch, Mercurio (2009) proposed a simple credit model, including the default risk relative to an average interbank counterparty

We assume that the risky Libor rate $L_x(T_1, T_2)$ fixed on the interbank market by the Libor panel is precisely the funding rate over the time interval $[T_1, T_2]$ of an abstract "average" Libor Bank. We may define as usual a discount factor $P_x(T_1, T_2)$ such that

$$P_x(T_1, T_2) = \frac{1}{1 + L_x(T_1, T_2)\tau(T_1, T_2)}. \tag{5}$$

This discount factor may be naturally associated with a risky zero coupon bond $P_x(T_1, T_2)$ issued by such average Libor Bank with maturity $T$. Denoting by $\tau_x(t)$ the default time at time $t$ of the Libor Bank, by $LGD_x$, $R_x = 1 - LGD_x$ the constant loss given default and recovery rate, respectively, associated with its default, and assuming independence between default and interest rates, we can price this zero coupon bond as

$$\begin{aligned}
P_x(t,T) &= E_t^{Q_d}\{D_d(t,T)[1_{[\tau_x(t)>T]} + R_x 1_{[\tau_x(t)\leq T]}]\} \\
&= E_t^{Q_d}\{D_d(t,T)[1 - 1_{[\tau_x(t)\leq T]} + R_x 1_{[\tau_x(t)\leq T]}]\} \\
&= E_t^{Q_d}\{D_d(t,T)[1 + LDG_x 1_{[\tau_x(t)\leq T]}]\} \\
&= E_t^{Q_d}[D_d(t,T)] + LDG_x E_t^{Q_d}[D_d(t,T)]E_t^{Q_d}[1_{[\tau_x(t)\leq T]}] \\
&= P_d(t,T) + LDG_x P_d(t,T)Q_{x,d}(t,T) \\
&= P_d(t,T)R(t,t,T,R_x),
\end{aligned}$$

$$D_d(t,T) = \exp\left[-\int_t^T r_d(u)du\right],$$

$$R(t,T_1,T_2,R_x) = 1 + LDG_x Q_{x,d}(t,T_1,T_2),$$

$$Q_{x,d}(t,T_1,T_2) = E_t^{Q_d}[Q_{x,d}(T_1,T_2)],$$

$$Q_{x,d}(T_1,T_2) = E_{T_1}^{Q_d}[1_{[\tau_x(T_1)\leq T_2]}],$$
(6)

where $t \leq T_1 < T_2$, $P_d(t.T)$ is the value in $t$ of a default free zero coupon bond with maturity $T$, $E_t^{Q_d}$ is the expected value in $t$ under the risk neutral probability measure $Q_d$, $r_d$ is the default free instantaneous interest rate, and $Q_{x,d}(T_1,T_2)$, $Q_{x,d}(t,T_1,T_2)$ are the spot and forward default probabilities of the Libor Bank, respectively. By considering the above assumptions, the risky Libor rate $L_x(T_1,T_2)$ is given by

$$L_x(T_1,T_2) = \frac{1}{\tau(T_1,T_2)}\left[\frac{1}{P_x(T_1,T_2)} - 1\right] = \frac{1}{\tau(T_1,T_2)}\left[\frac{1}{P_d(T_1,T_2)R(T_1,T_1,T_2,R_x)} - 1\right]. \quad (7)$$

Using equation 7 above, we can obtain the price in $t$ of a standard FRA contract that exchanges in $T_2$ the fixed rate $K$ versus the risky Libor rate $L_x(T_1,T_2)$ as

$$\begin{aligned}
FRA_{Std}(t,T_1,T_2,K) &= N\omega E_t^{Q_d}\{D_d(t,T_2)[L_x(T_1,T_2) - K]\tau(T_1,T_2)\} \\
&= N\omega E_t^{Q_d}\left\{D_d(t,T_1)D_d(T_1,T_2)\left[\frac{1}{P_x(T_1,T_2)} - 1 - K\tau(T_1,T_2)\right]\right\} \\
&= N\omega E_t^{Q_d}\left\{D_d(t,T_1)E_{T_1}^{Q_d}\left\{P_d(T_1,T_2)\left[\frac{1}{P_d(T_1,T_2)R(T_1,T_1,T_2,R_x)} - 1 - K\tau(T_1,T_2)\right]\right\}\right\} \\
&= N\omega E_t^{Q_d}\left\{D_d(t,T_1)E_{T_1}^{Q_d}\left\{\frac{1}{R(T_1,T_1,T_2,R_x)} - P_d(T_1,T_2)[1 + K\tau(T_1,T_2)]\right\}\right\} \\
&= N\omega E_t^{Q_d}\left\{\frac{D_d(t,T_1)}{R(T_1,T_1,T_2,R_x)} - D_d(t,T_2)[1 + K\tau(T_1,T_2)]\right\} \\
&= N\omega\left\{\frac{P_d(t,T_1)}{R(t,T_1,T_2,R_x)} - P_d(t,T_2)[1 + K\tau(T_1,T_2)]\right\},
\end{aligned}$$
(8)

where $N$ is the notional of the contract, $\omega = +1/-1$ for a payer/receiver FRA (referred to the fixed leg). The price of the market FRA is obtained through an analogous proof as

$$\begin{aligned}
FRA_{Mkt}(t,T_1,T_2,K) &= N\omega E_t^{Q_d}\left\{D_d(t,T_1)\frac{L_x(T_1,T_2) - K}{1 + L_x(T_1,T_2)\tau(T_1,T_2)}\tau(T_1,T_2)\right\} \\
&= N\omega\{P_d(t,T_1) - P_d(t,T_2)[1 + K\tau(T_1,T_2)]R(t,T_1,T_2,R_x)\} \\
&= FRA_{Std}(t,T_1,T_2,K)R(t,T_1,T_2,R_x).
\end{aligned}$$
(9)

We stress that the prices in equations 8 and 9 above have been nder the assumption that the FRA contract (not the underlying Libor rate) is credit risk free. Otherwise the derivation would involve the default indicator of the two counterparties involved in the FRA contract (not that of the average Libor Bank).

Assuming that the FRA contract is in equilibrium, such that $FRA(t,T_1,T_2,K) = 0$, denoting with $R^{FRA}(t,T_1,T_2) = K$ the equilibrium FRA rate at time $t$ and rearranging equations 8 and 9, we obtain

$$R^{FRA}_{Std}(t,T_1,T_2) = R^{FRA}_{Mkt}(t,T_1,T_2) = \frac{1}{\tau(T_1,T_2)}\left[\frac{P_d(t,T_1)}{P_d(t,T_2)}\frac{1}{R(t,T_1,T_2,R_x)} - 1\right]. \quad (10)$$

Since $0 \leq R_x \leq 1$ and $0 \leq Q_{x,d}(t,T_1,T_2) \leq 1$, we have that $0 \leq R(t,T_1,T_2,R_f) \leq 1$ then

$$\begin{aligned} R^{FRA}(t;T_1,T_2) &= \frac{1}{\tau(T_1,T_2)}\left[\frac{P_d(t,T_1)}{P_d(t,T_2)}\frac{1}{R(t,T_1,T_2,R_x)} - 1\right] \\ &\geq \frac{1}{\tau(T_1,T_2)}\left[\frac{P_d(t,T_1)}{P_d(t,T_2)} - 1\right] \\ &= F_d(t;T_1,T_2) \\ &= E_t^{Q_d^{T_2}}[L_d(T_1,T_2)] \end{aligned} \quad (11)$$

where $F_d(t,T_1,T_2)$ is the default free forward rate. We conclude that, thanks to default risk, the risky FRA rate is always higher than the corresponding forward rate relative to a default free yield curve. In other words, the default risk of the average Libor Bank, included into the Libor rate underlying a risk free FRA contract, induces a positive basis spread between the equilibrium FRA rate and a corresponding risk free forward rate. Only in the special case of a risk free Libor Bank, such that $Q_{x,d}(t,T_1,T_2) = 0$, we have

$$R(t,T_1,T_2,R_x) = 1 \;\forall\; t,T_1,T_2$$

$$R^{FRA}(t,T_1,T_2) = \frac{1}{\tau(T_1,T_2)}\left[\frac{P_d(t,T_1)}{P_d(t,T_2)} - 1\right] = F_d(t,T_1,T_2). \quad (12)$$

A risk free derivative could sound a little strange, in a market where even Libor Banks may default, but actually this ideal condition can be approximated in practice using collateralization, as discussed in the next section.

## 3. Collateralization and CSA Discounting

Another effect of the credit crunch has been the great diffusion of collateral agreements to reduce the counterparty risk of OTC derivatives positions. In the following sections we discuss some of the relevant aspects that concern the collateralization process.

### 3.1. Collateral Diffusion

Nowadays most of the counterparties on the interbank market have mutual collateral agreements in place. In 2011, almost 85% of all OTC derivatives transactions were collateralized, according to the ISDA (International Swaps and Derivatives Association) Margin Survey (ISDA 2012a). Respondents to the ISDA Margin Survey are divided in three different categories: large, medium and small dealers. The definition between the types of dealers is based on the number of active collateral agreements. Large dealers must have more than 3000 active agreements, while respondents that present active collateral agreements between 3000 and 100 are classified as medium market participant. Small dealers are financial institution and companies that report less than 100 active contracts. Among the 51 respondents, 14 are classified as large dealers and belong to the financial sector. The total number of respondents to the Margin Survey in 2012 is 51.

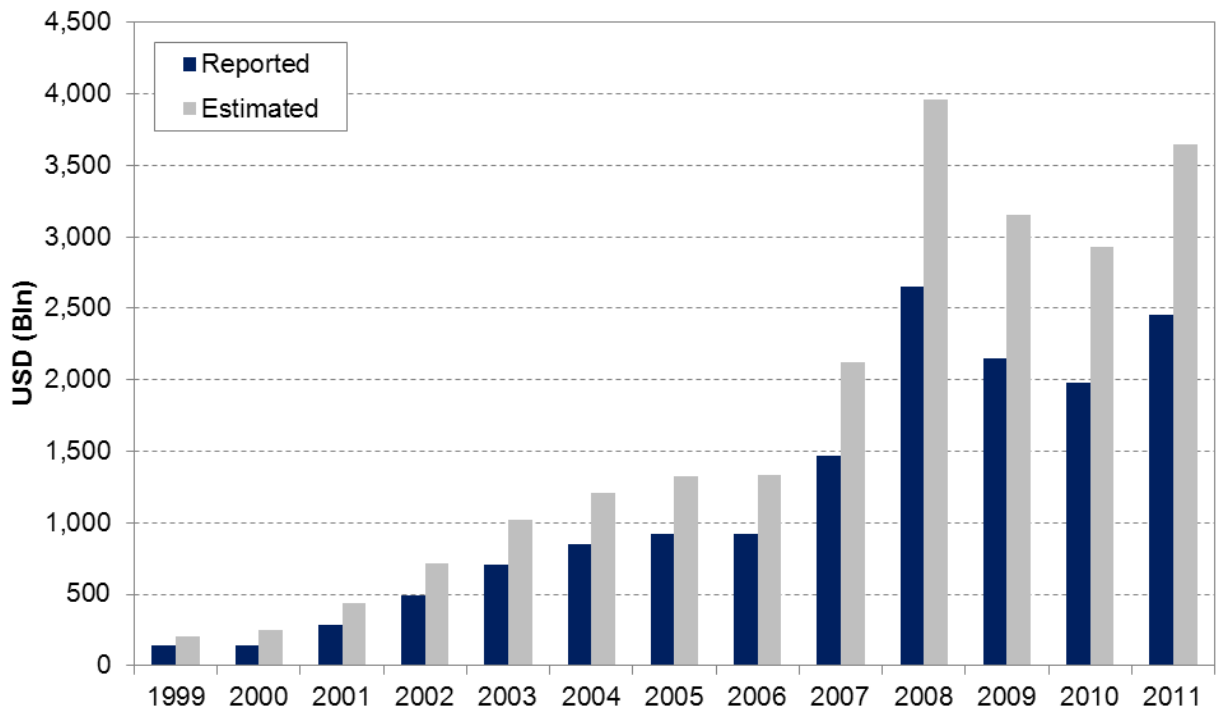

**Figure 13:** growth of the collateral value over the last 12 years (source: ISDA 2012a).

Figure 13 shows the growth in value of reported and estimated collateral in circulation within the OTC derivatives market. We can observe an upward trend of the collateral value over all the past 12 years, except between 2008 and 2010 where the reduction can be ascribed to a decrease in the market activity. During the 2011 the reported and the estimated value of collateral have experienced an increase of 24%.

As shown in Table 2, more than 80% of all OTC derivatives are collateralized among large dealers. Credit derivative contracts present the highest frequency of collateralization followed by Fixed Income derivatives. The FX derivatives show the lowest percentage of collateralization (58.3%), but their poor result can be explained by the fact that this type of contract is usually characterized by short maturities that, compared to the other contract classes, help to mitigate the counterparty risk. The percentages of collateralization related to large dealers are always higher than those of other classes of market participants, regardless the type of OTC Derivatives.

|                          | All     | Large Dealers |
|--------------------------|---------|---------------|
| All OTC Derivatives      | 71.4%   | 83.7%         |
| Fixed Income Derivatives | 78.1%   | 89.9%         |
| Credit Derivatives       | 93.4%   | 96.1%         |
| FX Derivatives           | 55.6%   | 70.6%         |
| Equity Derivatives       | 72.7%   | 85.3%         |
| Commodities Derivatives  | 56.3%   | 63.9%         |

**Table 2:** percentage of trades subjected to collateral agreements by derivative and dealer type (source: ISDA 2012a).

## 3.2. Collateral Mechanics

A typical financial transaction generates streams of future cash flows whose total net present value (NPV) is the algebraic sum of all discounted expected cash flows. Generally, each transaction implies a mutual credit exposure between two counterparties, let's say, a bank ($B$) and a generic market counterparty ($C$). If, at any time $t$ during the life of the transaction, for counterparty $B$ we have that $NPV_B(t) > 0$, then counterparty $B$ expects to receive, on average, future cash flows from counterparty $C$ (in other words, $B$ has an expected credit with $C$ at time $t$). On the other side, if counterparty $C$ has $NPV_C(t) < 0$, then it expects to pay, on average, future cash flows to counterparty $B$ (in other words, $C$ has an expected debt with $B$ at time $t$). The reverse holds if $NPV_B(t) < 0$ and $NPV_C(t) > 0$.

Such credit/debit exposure is clearly subject to bilateral default risk, the probability that the debtor counterparty may default, not fulfilling its obligations with the creditor counterparty. This credit risk can be mitigated through a guarantee, called "collateral agreement". This guarantee is formalised into an optional annex of the ISDA Master Agreement (the standard legal contract widely used to regulate OTC transactions) called Credit Support Annex (CSA). The main feature of the CSA is the additional obligation of the counterparties for a margination mechanism similar to those adopted by Central Counterparties (CCPs, i.e. LCH.Clearnet, Euroclear, SIX Securities Services, etc.) for OTC derivatives clearing, or by exchanges for standard market instruments clearing (i.e. Futures). In a nutshell, at every margination date the two counterparties check the value of the portfolio of mutual OTC transactions and regulate the margin, adding to or subtracting from the collateral account the corresponding mark to market variation with respect to the preceding margination date. The margination can be regulated with cash or with (primary) assets of corresponding value. In any case the collateral account holds, at each date, the total NPV of the portfolio, which is positive for the creditor counterparty and negative for the debtor counterparty. The collateral amount is available to the creditor.

On the other side, the debtor receives an interest on the collateral amount, called "collateral rate". Hence, we can look at the CSA as a funding mechanism, transferring liquidity from the debtor to the creditor. The main differences with traditional funding through Deposit contracts are that, using derivatives, we have longer maturities and variable (stochastic) lending/borrowing sides and amounts. We can also look at CSA as an hedging mechanism, where the collateral amount hedges the creditor against the event of default of the debtor. The most diffused CSA provides a daily margination mechanism and an overnight collateral rate (ISDA 2012a). Actual CSAs provide many other detailed features (i.e. credit support amount, delivery amount, minimum transfer amount, collateral currency, etc.) that are out of the scope of the present discussion.

Figure 14 illustrates a general scheme of funding with or without CSA that we briefly introduced above. When both the counterparties can post and receive collateral, the CSA mechanisms is called "two-way-CSA" (Figure 14, left panel) and it allows counterparties to fund OTC deals at the relevant collateral rate. If no collateral agreement is in place, the bank has to recur to the money market to find the necessary amounts $B(t)$ to fund the transaction, at the relevant money market rate (i.e. Euribor) plus, generally, a spread $\Delta(t)$ corresponding to its credit quality and interbank market conditions (Figure 14, right panel).

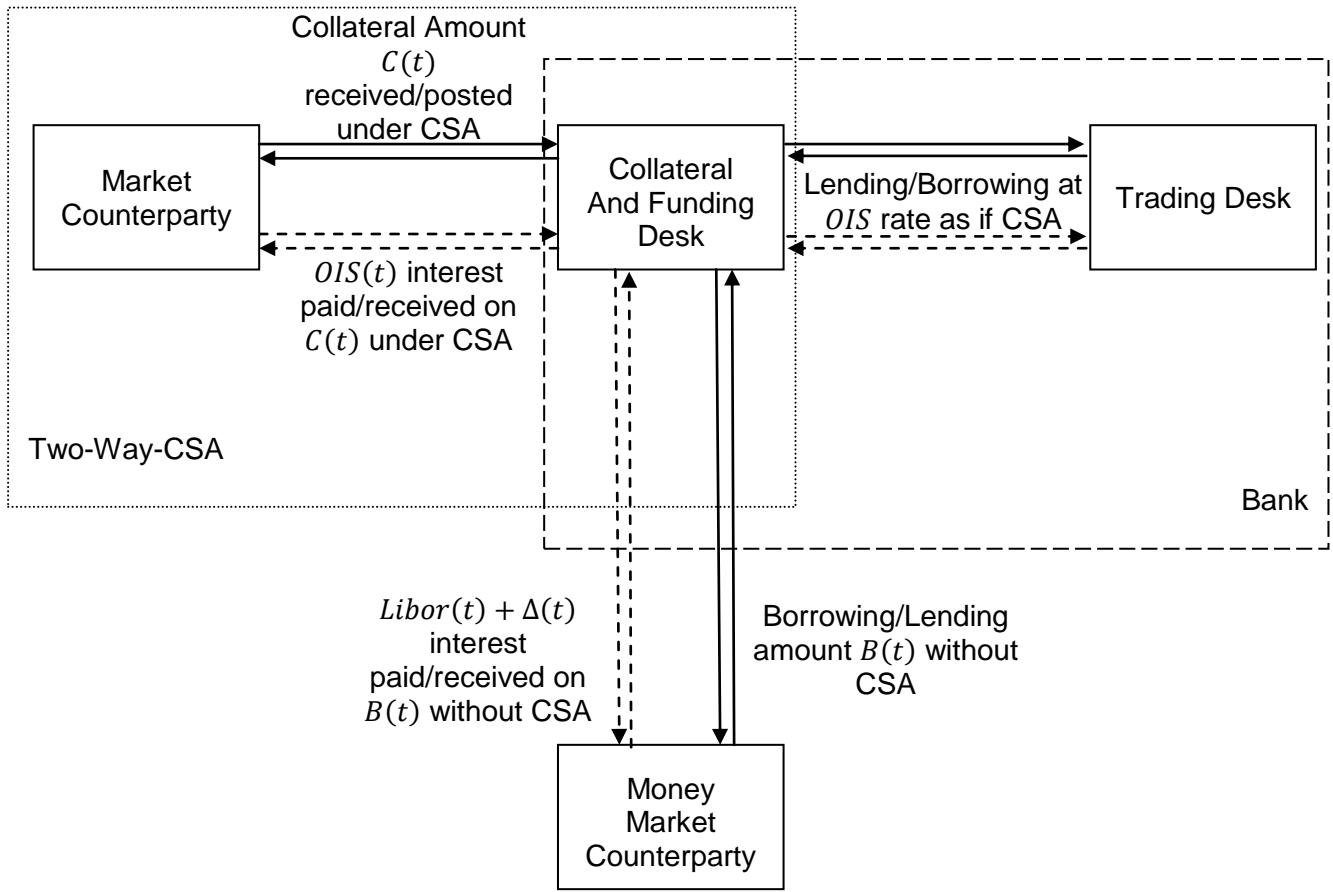

**Figure 14:** two-way-CSA with cash collateral and no-CSA funding mechanics.

## 3.3. CSA Discounting

A first important consequence of the diffusion of collateral agreements among interbank counterparties is that we can consider the derivatives' prices quoted on the OTC interbank market as transactions between counterparties under CSA. A second important consequence is that, by no-arbitrage and self-financing conditions, the CSA margination rate and the discounting rate of future cash flows must match, hence the name of "CSA discounting" (or "OIS discounting"). In particular, the most diffused overnight CSA implies overnight-based discounting and the construction of a discounting yield curve that must reflect, for each maturity, the funding level in an overnight collateralized interbank market. Thus OIS are the natural instruments for the discounting curve construction, that are also the best available proxies of risk free rates (see Ametrano and Bianchetti 2009, Bianchetti 2010).

The CSA discounting approach for the evaluation of collateralized OTC trades can be described by considering a simple cash flow transaction trade between two generic default free counterparties (bank $B$ and a generic market counterparty $C$) with payoff $\Pi(T)$ at maturity time $T$. We assume that the deal is under perfect collateralization with margination dates $\boldsymbol{T} = \{T_0, \ldots, T_i, \ldots, T_n\}$, where $T_n = T$, and perfect collateral account $B_c$ such that

$$B_c(t) = \Pi(t) \text{ for each } T_i \in \boldsymbol{T},$$
$$B_c(T_i) = B_c(T_{i-1})[1 + R_c(T_{i-1}) \Delta T_i],$$

where $R_c(T_{i-1})$ is the simply compounded collateral rate (i.e. Eonia) fixed at time $T_{i-1}$ and covering the interval $\Delta T_i$. The payoff of the trade and the structure of the margination dates are shown in Figure 15.

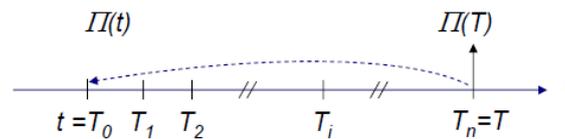

**Figure 15:** simple deal with a single cash flow payoff and multiple margination dates.

Suppose that the trade is under collateral with only two margination dates, $T_0$ and $T$, and $B$ receives a positive amount $\Pi(T)$ at maturity $T$, which correspond to a present value $\Pi(T_0)$ at time $T_0$, as depicted in Figure 16. Hence, counterparty $C$ posts an amount $B_c(T_0)$ into the collateral bank account that grows at the collateral rate $R_c(T_0)$ up to maturity $T$. By no-arbitrage and self-financing we have

$$B_c(T) = B_c(T_0)[1 + R_c(T_0)(T - T_0)] = \Pi(T),$$

$$\Pi(T_0) = P_d(T_0, T)\Pi(T) = B_c(T_0) \text{ where } P_d(T_0, T) = \frac{1}{[1 + R_c(T_0)(T - T_0)]}.$$

We can conclude that, by simple no-arbitrage arguments, future cash flows associated with collateralized trades must be discounted at the collateral rate.

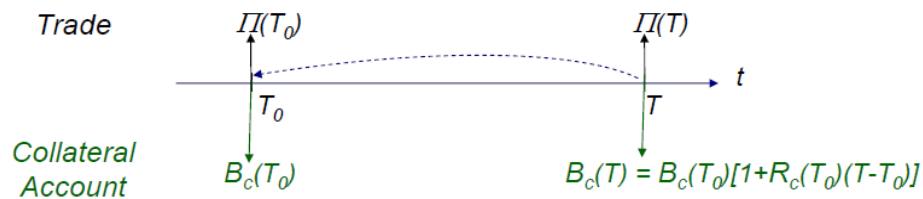

**Figure 16:** trade and collateral account cash flows scheme.

In case of absence of CSA, using the same no-arbitrage and self-financing principles between the funding and the discounting rate, we argue that future cash flows (positive or negative) must be discounted using the corresponding funding term structure. This implies important (and rather involved) consequences, such that, according with Morini and Prampolini (2011), each counterparty assigns a different present value to the same future cash flow, breaking the fair value symmetry. Indeed, a worsening of the its credit standing allows the Bank to sell derivatives (options in particular) at more competitive prices (the lower the rate, the higher the discount, the lower the price).

We stress that before the crisis the old-style standard Libor curve was representative of the average funding level on the interbank market (see e.g. Hull 2008). Such curve, even if considered a good proxy for a risk free curve, due to the perceived low counterparty risk of primary banks (belonging to the Libor Contribution panel), was not strictly risk free because of the absence of collateralization.

Even if the collateralization of OTC contracts has become the best market practice since the start of the financial crisis, there are still some confusion and controversial aspects (Sawyer 2011a, 2011b, 2012). For example, most of the CSAs allow collateral in multiple eligible types (e.g. cash, securities, or assets) and currencies (e.g. EUR, USD). Counterparties are often allowed to switch collateral during the life of the deal. Following the discussion above, changing e.g. the collateral currency implies changing the discounting curve, that leads to a change in the NPV of the contract. Hence, multi-currency CSA implies a cheapest-to-deliver collateral currency option on the underlying portfolio of instruments under CSA. This is an important friction against unwinding and back-loading existing trades on to CCPs because changes in the discounting curves and collateral currency options may be very expensive.

In order to avoid confusion in the market, reduce margin disputes and settlement risk (i.e. Herstatt risk), ISDA has developed a new Standard CSA (SCSA) where the hidden optionalities of the old

CSA are reduced. The new SCSA introduces different "silos" corresponding to the currencies with the most liquid OIS curves (e.g. EUR, USD, GBP, CHF, JPY). Each transaction is allocated to one single silo, in relation with the currency of the underlying, with the relevant OIS rate used to discount all the trade's cash flows. According to this mechanics, counterparties will have to manage multiple flows of collateral in different currencies that will create a systemic cross currency settlement risk (i.e. two counterparties, two swaps, one in EUR, one in USD). To solve the problem the SCSA allows counterparties to net collateral flows denominated in multiple currencies into a single payment in a single currency using the overnight currency swap market to exchange the flows (ISDA 2012b). However, implementing the SCSA requires to solve some relevant challenges such as the adoption of a common set of exchange rates and calculation specifications for managing collateralized trades.

## 3.4. Market Quotes

During the financial crisis, the market has experienced a transitional phase from the classical Libor-based discounting methodology to the modern CSA based discounting methodology. As we notice in section 3.1, OTC transactions executed on the interbank market normally use CSA discounting. In particular, plain vanilla interest rate derivatives, such as FRA, Swaps, Basis Swaps, Caps/Floor/Swaptions are quoted by main brokers using CSA discounting.

In particular in September 2010 the international broker ICAP switched to OIS discounting, publishing both Libor- and OIS-based cap/floor/swaption implied volatilities. Since January 2012 only OIS-based volatilities are published (ICAP 2010, 2011).

In order to appreciate the implied evolution of the market quotes, we show in Figure 17, Figure 18, Figure 19 below the market quotations of EUR Swaption premia and volatilities on three different dates that cover the period Jun. 2010 – May 2012. Figure 17 shows the standard pre-crisis quotes for EUR Swaption premia and volatilities (reference date 30 June 2010), obtained by using the classical single-curve approach. Figure 18 shows the quotes for EUR Swaption on 30 September 2010, where we can notice two new panels, Figure 18 – panel C and Figure 18 – panel D. The first shows the forward EUR Swaption premia, while the second shows the implied EUR ATM Swaption volatility surface obtained using the modern multiple-curve CSA discounting methodology (i.e. Eonia discounts, Euribor6M forwards, consistent with EUR Swaps and CSA). The implied volatility surface of Figure 18 – panel D is retrieved from the quoted premia using classical Euribor discounting. We point out that the two different implied volatility surfaces, coherently with the two pricing methodologies, lead to the same premia, assuming the instrument is traded between OTC counterparties under mutual SCSA in EUR . Finally, Figure 19 shows the market quotes for EUR Swaption on 31 January 2012. Only the OIS-based implied volatility surface appears (Figure 19 – panel D), while both Euribor- and Eonia-based spot prices are reported (Figure 19 – panel B and C respectively), consistently with the two different pricing approaches for trades under collateral or not.

| A | 30/06/2010 - EUR ATM Swaption Straddles - Premium Mids | | | | | | | | | | | | | | B | 30/06/2010 - EUR ATM Swaption Straddles - Implied Volatilities | | | | | | | | | | | | | |
|---|---|---|---|---|---|---|---|---|---|---|---|---|---|---|---|---|---|---|---|---|---|---|---|---|---|---|---|---|
| | Swap Tenor | | | | | | | | | | | | | | | Maturity | | | | | | | | | | | | | |
| Option Expiry | 1Y | 2Y | 3Y | 4Y | 5Y | 6Y | 7Y | 8Y | 9Y | 10Y | 15Y | 20Y | 25Y | 30Y | Option Expiry | 1Y | 2Y | 3Y | 4Y | 5Y | 6Y | 7Y | 8Y | 9Y | 10Y | 15Y | 20Y | 25Y | 30Y |
| 1M Opt | 11 | 24.5 | 38.5 | 53 | 68 | 83.5 | 99 | 115 | 129 | 144 | 205 | 261 | 319 | 389 | 1M Opt | 46.4 | 39.2 | 35.9 | 32.9 | 29.9 | 28.1 | 27 | 26.3 | 25.1 | 23.3 | 23.5 | 24.9 | 27.6 |
| 2M Opt | 16.5 | 37 | 58 | 79.5 | 101 | 122 | 143 | 163 | 183 | 203 | 290 | 373 | 456 | 548 | 2M Opt | 48.8 | 40.6 | 37 | 33.4 | 30.8 | 28.6 | 27.1 | 25.9 | 25.2 | 24.7 | 23.1 | 23.5 | 25 | 27.3 |
| 3M Opt | 22.5 | 50.5 | 77.5 | 104 | 129 | 153 | 177 | 201 | 225 | 250 | 355 | 458 | 564 | 666 | 3M Opt | 51.7 | 44.1 | 39.5 | 34.8 | 31.2 | 28.7 | 27 | 25 | 24.6 | 23 | 23.4 | 25.1 | 27 |
| 6M Opt | 37.5 | 81 | 121 | 158 | 195 | 231 | 265 | 298 | 330 | 362 | 499 | 624 | 763 | 919 | 6M Opt | 57.9 | 47.4 | 41.2 | 35.9 | 32.2 | 29.7 | 27.8 | 26.5 | 25.5 | 24.8 | 22.7 | 24.5 | 24.1 | 26.4 |
| 9M Opt | 50 | 107 | 157 | 205 | 252 | 295 | 337 | 378 | 417 | 455 | 622 | 767 | 931 | 1104 | 9M Opt | 59.9 | 48.4 | 41.4 | 36.2 | 32.7 | 30.1 | 28.2 | 26.9 | 25.9 | 25.1 | 22.9 | 22.6 | 24 | 26 |
| 1Y Opt | 61.5 | 126 | 186 | 245 | 300 | 352 | 402 | 449 | 495 | 536 | 734 | 900 | 1081 | 1256 | 1Y Opt | 59.5 | 46.6 | 40.2 | 35.7 | 32.4 | 30 | 28.3 | 27.1 | 26.1 | 25.1 | 23.2 | 22.9 | 23.3 | 25.6 |
| 18M Opt | 78.5 | 158 | 233 | 305 | 373 | 437 | 498 | 557 | 612 | 666 | 902 | 1116 | 1318 | 1512 | 18M Opt | 53.9 | 42.1 | 36.9 | 33.2 | 30.6 | 28.7 | 27.4 | 26.3 | 25.4 | 24.8 | 23 | 23.1 | 24 | 25.2 |
| 2Y Opt | 94 | 183 | 268 | 349 | 427 | 503 | 574 | 641 | 705 | 767 | 1028 | 1259 | 1492 | 1708 | 2Y Opt | 48.2 | 37.6 | 33.5 | 30.6 | 28.6 | 27.3 | 26.3 | 25.4 | 24.6 | 24.1 | 22.5 | 22.5 | 23.6 | 24.7 |
| 3Y Opt | 111 | 214 | 312 | 405 | 494 | 581 | 663 | 740 | 815 | 888 | 1182 | 1447 | 1726 | 1991 | 3Y Opt | 35.5 | 29.9 | 27.4 | 25.8 | 24.7 | 24 | 23.3 | 22.8 | 22.3 | 22 | 20.9 | 21.2 | 22.5 | 23.8 |
| 4Y Opt | 121.5 | 235 | 341 | 441 | 537 | 630 | 719 | 803 | 885 | 965 | 1282 | 1571 | 1874 | 2153 | 4Y Opt | 28.3 | 25.2 | 23.8 | 22.9 | 22.2 | 21.7 | 21.2 | 20.9 | 20.6 | 20.5 | 19.8 | 20.3 | 21.6 | 22.8 |
| 5Y Opt | 126.5 | 244 | 357 | 464 | 565 | 661 | 754 | 842 | 929 | 1011 | 1351 | 1656 | 1963 | 2274 | 5Y Opt | 24 | 22.1 | 21.5 | 21 | 20.5 | 20.1 | 19.8 | 19.6 | 19.4 | 19.4 | 19.1 | 19.7 | 21 | 22.3 |
| 7Y Opt | 129 | 250 | 367 | 480 | 587 | 686 | 783 | 878 | 971 | 1062 | 1411 | 1730 | 2049 | 2375 | 7Y Opt | 19.7 | 19 | 18.8 | 18.5 | 18.3 | 18.1 | 18 | 18 | 18.5 | 18.1 | 18 | 18.2 | 18.2 | 19 | 20.2 | 21.4 |
| 10Y Opt | 128.5 | 247 | 364 | 477 | 586 | 689 | 791 | 891 | 988 | 1082 | 1444 | 1763 | 2092 | 2404 | 10Y Opt | 16.9 | 16.5 | 16.5 | 16.6 | 16.8 | 16.9 | 17 | 17.2 | 17.5 | 17.7 | 18.2 | 19.1 | 20.2 | 21.2 |
| 15Y Opt | 123 | 235 | 348 | 461 | 569 | 673 | 775 | 874 | 968 | 1057 | 1407 | 1722 | 2043 | 2352 | 15Y Opt | 17 | 16.9 | 17.2 | 17.6 | 18.1 | 18.5 | 18.8 | 19.3 | 19.7 | 20 | 20.7 | 21.3 | 22 | 22.6 |
| 20Y Opt | 113.5 | 220 | 326 | 434 | 540 | 640 | 736 | 829 | 915 | 994 | 1337 | 1627 | 1910 | 2164 | 20Y Opt | 19.9 | 20.2 | 20.8 | 21.8 | 22.6 | 23.2 | 23.6 | 24 | 24.3 | 24.6 | 24.7 | 24.4 | 23.8 | 23.3 |
| 25Y Opt | 104.5 | 201 | 300 | 401 | 500 | 591 | 676 | 764 | 843 | 918 | 1220 | 1483 | 1726 | 1946 | 25Y Opt | 26 | 26.1 | 26.8 | 27.8 | 28.5 | 28.6 | 28.4 | 28.4 | 28 | 27.7 | 26.5 | 25.3 | 23.7 | 22.7 | 22.4 |
| 30Y Opt | 95 | 180 | 267 | 355 | 442 | 522 | 598 | 676 | 746 | 812 | 1092 | 1337 | 1569 | 1791 | 30Y Opt | 27.6 | 26.2 | 26 | 26 | 26.1 | 25.8 | 25.5 | 25.4 | 25 | 24.6 | 22.4 | 21.1 | 20.8 | 21.2 |

**Figure 17:** EUR at-the-money Swaption market quotes on 30 June 2010. Premia (left panel) and Black's implied volatilities (right panel). The ATM implied volatilities surface is obtained using the classical single-curve approach. (Source: ICAP).

| A | 30/09/2010 - EUR ATM Swaption Straddles - Implied Spot Premium Mids | | | | | | | | | | | | | | B | 30/09/2010 - EUR ATM Swaption Straddles - Implied Volatilities (Euribor disc) | | | | | | | | | | | | | |
|---|---|---|---|---|---|---|---|---|---|---|---|---|---|---|---|---|---|---|---|---|---|---|---|---|---|---|---|---|
| | Swap Tenor | | | | | | | | | | | | | | | | | | | | | | | | | | | | |
| Option Expiry | 1Y | 2Y | 3Y | 4Y | 5Y | 6Y | 7Y | 8Y | 9Y | 10Y | 15Y | 20Y | 25Y | 30Y | Option Expiry | 1Y | 2Y | 3Y | 4Y | 5Y | 6Y | 7Y | 8Y | 9Y | 10Y | 15Y | 20Y | 25Y | 30Y |
| 1M Opt | 10.5 | 25.5 | 41 | 57.5 | 76.5 | 96.5 | 117 | 138 | 161 | 185 | 270 | 346 | 420 | 494 | 1M Opt | 40.5 | 38.4 | 37.9 | 36.8 | 36.3 | 35.9 | 35.5 | 35.4 | 35.8 | 36.4 | 34.4 | 34.4 | 36.1 | 38.6 |
| 2M Opt | 15.5 | 37.5 | 60 | 85.5 | 115 | 141 | 168 | 198 | 227 | 261 | 379 | 486 | 592 | 700 | 2M Opt | 40.4 | 38.6 | 37.6 | 37.2 | 37.1 | 35.8 | 34.9 | 34.7 | 34.6 | 35.1 | 33.2 | 33.3 | 35 | 37.7 |
| 3M Opt | 20 | 49.5 | 78.5 | 110 | 142 | 176 | 209 | 243 | 278 | 315 | 458 | 583 | 711 | 841 | 3M Opt | 41.1 | 40.8 | 39.5 | 38.4 | 37.2 | 36.1 | 35.1 | 34.6 | 34.4 | 34.5 | 32.8 | 32.7 | 34.5 | 37.1 |
| 6M Opt | 32 | 76.5 | 116 | 159 | 205 | 250 | 293 | 340 | 385 | 430 | 620 | 796 | 961 | 1131 | 6M Opt | 44.1 | 42.8 | 39.8 | 38.1 | 36.8 | 35.3 | 34.2 | 33.7 | 33.2 | 32.9 | 31.2 | 31.6 | 33 | 35.4 |
| 9M Opt | 44.5 | 98.5 | 149 | 202 | 255 | 308 | 361 | 414 | 468 | 523 | 751 | 953 | 1148 | 1358 | 9M Opt | 47 | 43.1 | 40.3 | 38 | 36.1 | 34.6 | 33.5 | 32.8 | 32.4 | 32.1 | 30.6 | 30.7 | 32.1 | 34.6 |
| 1Y Opt | 54.5 | 117 | 177 | 235 | 296 | 360 | 419 | 479 | 540 | 599 | 852 | 1074 | 1300 | 1526 | 1Y Opt | 47.6 | 42.7 | 39.7 | 37 | 35.2 | 34.1 | 33 | 32.3 | 31.4 | 29.9 | 29.9 | 31.5 | 33.7 |
| 18M Opt | 75 | 149 | 222 | 296 | 369 | 441 | 512 | 582 | 655 | 729 | 1020 | 1286 | 1542 | 1792 | 18M Opt | 48.8 | 41.2 | 38.1 | 35.8 | 34 | 32.7 | 31.8 | 31.1 | 30.7 | 30.5 | 29 | 29.2 | 30.6 | 32.5 |
| 2Y Opt | 92.5 | 179 | 262 | 344 | 428 | 510 | 589 | 667 | 747 | 825 | 1156 | 1449 | 1731 | 2002 | 2Y Opt | 47.8 | 40.1 | 36.7 | 34.1 | 32.6 | 31.5 | 30.7 | 30 | 29.6 | 29.3 | 28.2 | 28.5 | 29.9 | 31.6 |
| 3Y Opt | 112.5 | 217 | 315 | 411 | 509 | 604 | 697 | 785 | 874 | 967 | 1328 | 1651 | 1992 | 2305 | 3Y Opt | 40.8 | 35 | 32.2 | 30.4 | 29.5 | 28.7 | 28.2 | 27.6 | 27.3 | 26.3 | 26.9 | 28.5 | 30.2 |
| 4Y Opt | 124.5 | 240 | 349 | 455 | 556 | 658 | 756 | 852 | 946 | 1049 | 1427 | 1778 | 2125 | 2447 | 4Y Opt | 34.3 | 30.3 | 28.5 | 27.4 | 26.6 | 26.1 | 25.6 | 25.3 | 25.1 | 25 | 24.5 | 25.3 | 26.9 | 28.4 |
| 5Y Opt | 131 | 253 | 367 | 476 | 583 | 689 | 792 | 895 | 996 | 1100 | 1497 | 1852 | 2197 | 2549 | 5Y Opt | 29.2 | 26.7 | 25.6 | 24.8 | 24.3 | 24 | 23.7 | 23.6 | 23.6 | 23.8 | 23.4 | 24.1 | 25.6 | 27.2 |
| 7Y Opt | 133.5 | 259 | 377 | 494 | 606 | 717 | 825 | 928 | 1031 | 1141 | 1539 | 1900 | 2249 | 2605 | 7Y Opt | 23.6 | 22.5 | 22 | 21.6 | 21.3 | 21.2 | 21.3 | 21.5 | 21.4 | 21.7 | 21.6 | 22.6 | 23.9 | 25.3 |
| 10Y Opt | 136 | 263 | 388 | 508 | 626 | 739 | 849 | 953 | 1058 | 1166 | 1563 | 1928 | 2270 | 2594 | 10Y Opt | 20 | 19.5 | 19.6 | 19.7 | 19.8 | 19.9 | 20 | 20.1 | 20.3 | 20.6 | 21.1 | 22.3 | 23.4 | 24.3 |
| 15Y Opt | 130 | 251 | 371 | 487 | 598 | 709 | 817 | 922 | 1029 | 1134 | 1515 | 1888 | 2223 | 2539 | 15Y Opt | 19.3 | 19.2 | 19.4 | 19.6 | 19.8 | 20.2 | 20.7 | 21.2 | 21.7 | 22.3 | 23.3 | 24.4 | 25.1 | 25.4 |
| 20Y Opt | 119.5 | 235 | 348 | 460 | 569 | 676 | 780 | 881 | 981 | 1068 | 1444 | 1780 | 2087 | 2361 | 20Y Opt | 21.4 | 22.1 | 22.9 | 23.7 | 24.5 | 25.3 | 26.1 | 26.8 | 27.4 | 27.7 | 27.8 | 27.5 | 27 | 26.3 |
| 25Y Opt | 110 | 214 | 320 | 426 | 528 | 628 | 722 | 811 | 904 | 990 | 1324 | 1621 | 1896 | 2142 | 25Y Opt | 28.2 | 29.3 | 30.3 | 31.4 | 32.1 | 32.6 | 32.6 | 32.5 | 32.6 | 32.3 | 29.6 | 27.5 | 26.1 | 25.5 |
| 30Y Opt | 101.5 | 196 | 289 | 384 | 477 | 565 | 651 | 732 | 816 | 893 | 1203 | 1485 | 1751 | 2012 | 30Y Opt | 32.5 | 31.3 | 31.1 | 31.1 | 31 | 30.9 | 30.7 | 30.3 | 30.1 | 29.6 | 26.2 | 24.4 | 23.9 | 24.4 |

| C | 30/09/2010 - EUR ATM Swaption Straddles - Fwd Premium Mids | | | | | | | | | | | | | | D | 30/09/2010 - EUR ATM Swaption Straddles - Implied Volatilities | | | | | | | | | | | | | |
|---|---|---|---|---|---|---|---|---|---|---|---|---|---|---|---|---|---|---|---|---|---|---|---|---|---|---|---|---|
| | Swap Tenor | | | | | | | | | | | | | | | Swap Tenor | | | | | | | | | | | | | |
| Option Expiry | 1Y | 2Y | 3Y | 4Y | 5Y | 6Y | 7Y | 8Y | 9Y | 10Y | 15Y | 20Y | 25Y | 30Y | Option Expiry | 1Y | 2Y | 3Y | 4Y | 5Y | 6Y | 7Y | 8Y | 9Y | 10Y | 15Y | 20Y | 25Y | 30Y |
| 1M Opt | 10.5 | 25.5 | 41 | 57.5 | 76.5 | 96.5 | 117 | 138 | 161 | 185 | 270 | 346 | 421 | 495 | 1M Opt | 40.6 | 38.4 | 37.9 | 36.8 | 36.3 | 35.8 | 35.5 | 35.4 | 35.8 | 36.4 | 34.4 | 34.4 | 36.1 | 38.6 |
| 2M Opt | 15.5 | 37.5 | 60 | 86 | 115 | 141 | 169 | 198 | 228 | 261 | 379 | 486 | 592 | 701 | 2M Opt | 40.4 | 38.6 | 37.6 | 37.1 | 37.1 | 35.8 | 34.9 | 34.7 | 34.6 | 35.1 | 33.2 | 33.3 | 35 | 37.7 |
| 3M Opt | 20 | 49.5 | 78.5 | 110 | 142 | 176 | 209 | 243 | 278 | 315 | 459 | 584 | 712 | 842 | 3M Opt | 41.1 | 40.8 | 39.5 | 38.3 | 37.1 | 36.1 | 35.1 | 34.6 | 34.4 | 34.5 | 32.8 | 32.7 | 34.4 | 37 |
| 6M Opt | 32 | 76.5 | 116 | 160 | 206 | 250 | 294 | 341 | 386 | 431 | 622 | 799 | 964 | 1135 | 6M Opt | 44.1 | 42.7 | 39.7 | 38 | 36.7 | 35.3 | 34.1 | 33.6 | 33.1 | 32.8 | 31.5 | 31.5 | 33 | 35.3 |
| 9M Opt | 44.5 | 99 | 150 | 203 | 257 | 310 | 363 | 417 | 471 | 526 | 756 | 959 | 1155 | 1366 | 9M Opt | 46.9 | 43 | 40.2 | 37.8 | 36 | 34.5 | 33.4 | 32.7 | 32.3 | 32.1 | 30.6 | 30.6 | 32.1 | 34.5 |
| 1Y Opt | 55 | 118 | 178 | 237 | 299 | 363 | 423 | 483 | 545 | 605 | 859 | 1083 | 1311 | 1539 | 1Y Opt | 47.5 | 42.5 | 39.6 | 36.8 | 35.1 | 34 | 32.9 | 32.2 | 31.7 | 31.3 | 29.8 | 29.8 | 31.4 | 33.6 |
| 18M Opt | 76 | 151 | 225 | 300 | 374 | 447 | 519 | 590 | 664 | 739 | 1035 | 1303 | 1564 | 1817 | 18M Opt | 48.7 | 41 | 37.9 | 35.6 | 33.8 | 32.5 | 31.6 | 30.9 | 30.5 | 30.3 | 28.9 | 29 | 30.5 | 32.3 |
| 2Y Opt | 94.5 | 183 | 268 | 351 | 437 | 520 | 601 | 681 | 762 | 842 | 1179 | 1479 | 1767 | 2042 | 2Y Opt | 47.6 | 39.8 | 36.4 | 33.8 | 32.4 | 31.3 | 30.5 | 29.8 | 29.4 | 29.1 | 28 | 28.3 | 29.7 | 31.4 |
| 3Y Opt | 116.5 | 225 | 327 | 426 | 528 | 625 | 722 | 814 | 906 | 1002 | 1376 | 1726 | 2064 | 2388 | 3Y Opt | 40.5 | 34.6 | 31.9 | 30 | 29.2 | 28.4 | 27.9 | 27.4 | 27 | 27 | 26 | 26.6 | 28.2 | 29.8 |
| 4Y Opt | 131.5 | 254 | 369 | 480 | 588 | 695 | 798 | 900 | 999 | 1108 | 1507 | 1878 | 2244 | 2585 | 4Y Opt | 34 | 29.9 | 28.1 | 27 | 26.2 | 25.7 | 25.3 | 25 | 24.8 | 24.9 | 24.2 | 24.9 | 26.5 | 27.9 |
| 5Y Opt | 141.5 | 273 | 396 | 514 | 630 | 744 | 856 | 967 | 1076 | 1189 | 1618 | 2001 | 2374 | 2754 | 5Y Opt | 28.9 | 26.3 | 25.2 | 24.4 | 23.9 | 23.6 | 23.3 | 23.2 | 23.2 | 23.4 | 23 | 23.7 | 25.1 | 26.7 |
| 7Y Opt | 152 | 295 | 431 | 564 | 691 | 819 | 942 | 1059 | 1177 | 1302 | 1757 | 2168 | 2567 | 2974 | 7Y Opt | 23.2 | 22 | 21.5 | 21.1 | 20.8 | 20.8 | 20.8 | 20.9 | 21.2 | 21.1 | 22 | 23.3 | 24.7 |
| 10Y Opt | 170.5 | 329 | 487 | 638 | 785 | 927 | 1065 | 1195 | 1327 | 1463 | 1960 | 2419 | 2848 | 3255 | 10Y Opt | 19.6 | 18.9 | 19 | 19.1 | 19.2 | 19.3 | 19.4 | 19.5 | 19.7 | 20 | 20.4 | 21.6 | 22.6 | 23.5 |
| 15Y Opt | 193 | 374 | 552 | 725 | 890 | 1055 | 1217 | 1373 | 1532 | 1688 | 2256 | 2810 | 3309 | 3780 | 15Y Opt | 18.8 | 18.4 | 18.6 | 18.8 | 19 | 19.4 | 19.8 | 20.3 | 20.8 | 21.4 | 22.2 | 23.3 | 24.2 |
| 20Y Opt | 209 | 410 | 608 | 804 | 994 | 1181 | 1363 | 1539 | 1714 | 1866 | 2523 | 3110 | 3645 | 4115 | 20Y Opt | 20.6 | 21.6 | 22.4 | 23.1 | 23.9 | 24.5 | 25.2 | 25.8 | 26 | 26 | 25.8 | 25.3 | 24.7 |
| 25Y Opt | 218 | 424 | 633 | 843 | 1047 | 1244 | 1430 | 1607 | 1791 | 1961 | 2622 | 3209 | 3755 | 4241 | 25Y Opt | 27.2 | 27 | 28.1 | 29 | 29.4 | 29.6 | 29.4 | 29.2 | 29.3 | 29.6 | 27.2 | 25.3 | 24.1 | 23.7 |
| 30Y Opt | 220.5 | 425 | 629 | 833 | 1035 | 1227 | 1414 | 1590 | 1771 | 1939 | 2613 | 3226 | 3803 | 4369 | 30Y Opt | 29.2 | 26.9 | 26.7 | 26.9 | 28 | 27.1 | 27.2 | 27 | 27.1 | 26.8 | 23.9 | 22.3 | 22 | 22.3 |

**Figure 18:** EUR at-the-money Swaption market quotes on 30 September 2010. Premia are distinguished between spot (panel A) and forward ones (panel C). Implied spot premia are obtained from forward premia using Eonia discounting. Implied volatilities quotes are now differentiated between the Euribor one (panel B) and the Eonia one (panel D). (Source: ICAP).

**A** 31/05/2012 - EUR ATM Swaption Straddles - Implied Spot Premium Mids

| Option Expiry | 1Y | 2Y | 3Y | 4Y | 5Y | 6Y | 7Y | 8Y | 9Y | 10Y | 15Y | 20Y | 25Y | 30Y |
|---|---|---|---|---|---|---|---|---|---|---|---|---|---|---|
| 1M Opt | 12 | 24.5 | 39 | 56 | 73.5 | 92 | 112 | 133 | 154 | 177 | 271 | 362 | 445 | 523 |
| 2M Opt | 16.5 | 34.5 | 56 | 78.5 | 106 | 132 | 161 | 189 | 218 | 249 | 379 | 510 | 627 | 734 |
| 3M Opt | 19.5 | 43 | 69 | 100 | 131 | 163 | 198 | 233 | 271 | 310 | 461 | 613 | 753 | 876 |
| 6M Opt | 27.5 | 60 | 98 | 143 | 192 | 240 | 287 | 335 | 385 | 435 | 639 | 833 | 1026 | 1205 |
| 9M Opt | 33 | 75 | 120 | 174 | 239 | 297 | 355 | 415 | 476 | 533 | 771 | 997 | 1219 | 1431 |
| 1Y Opt | 39 | 86.5 | 143 | 210 | 282 | 349 | 418 | 486 | 555 | 622 | 892 | 1157 | 1409 | 1643 |
| 18M Opt | 52 | 115 | 187 | 273 | 359 | 441 | 523 | 604 | 684 | 762 | 1084 | 1393 | 1687 | 1974 |
| 2Y Opt | 71 | 150 | 240 | 335 | 433 | 523 | 612 | 701 | 788 | 877 | 1247 | 1594 | 1925 | 2247 |
| 3Y Opt | 106.5 | 215 | 326 | 439 | 554 | 659 | 762 | 864 | 963 | 1064 | 1494 | 1902 | 2297 | 2674 |
| 4Y Opt | 132.5 | 260 | 387 | 515 | 640 | 758 | 874 | 986 | 1095 | 1203 | 1670 | 2119 | 2554 | 2965 |
| 5Y Opt | 150 | 292 | 429 | 566 | 700 | 829 | 954 | 1076 | 1195 | 1308 | 1796 | 2264 | 2723 | 3153 |
| 7Y Opt | 169 | 327 | 476 | 624 | 771 | 912 | 1051 | 1186 | 1320 | 1450 | 1958 | 2437 | 2905 | 3351 |
| 10Y Opt | 176.5 | 347 | 506 | 661 | 815 | 967 | 1117 | 1265 | 1409 | 1549 | 2071 | 2554 | 3008 | 3429 |
| 15Y Opt | 175 | 346 | 508 | 669 | 825 | 977 | 1127 | 1276 | 1421 | 1571 | 2075 | 2537 | 2926 | 3312 |
| 20Y Opt | 169.5 | 335 | 496 | 652 | 805 | 955 | 1102 | 1246 | 1387 | 1523 | 2013 | 2426 | 2816 | 3161 |
| 25Y Opt | 162 | 321 | 477 | 625 | 771 | 915 | 1057 | 1197 | 1335 | 1475 | 1931 | 2310 | 2683 | 2999 |
| 30Y Opt | 157.5 | 309 | 455 | 595 | 732 | 868 | 1003 | 1137 | 1273 | 1416 | 1847 | 2213 | 2562 | 2888 |

**B** 31/05/2012 - EUR ATM Swaption Straddles - Implied Spot Premium Mids (Euribor disc)

| Option Expiry | 1Y | 2Y | 3Y | 4Y | 5Y | 6Y | 7Y | 8Y | 9Y | 10Y | 15Y | 20Y | 25Y | 30Y |
|---|---|---|---|---|---|---|---|---|---|---|---|---|---|---|
| 1M Opt | 12 | 25 | 39 | 56 | 74 | 92 | 112 | 132 | 154 | 177 | 270 | 361 | 444 | 523 |
| 2M Opt | 16 | 34 | 56 | 78 | 105 | 132 | 161 | 188 | 217 | 248 | 378 | 509 | 626 | 733 |
| 3M Opt | 20 | 43 | 69 | 100 | 131 | 163 | 197 | 233 | 270 | 308 | 460 | 611 | 751 | 874 |
| 6M Opt | 27 | 59 | 97 | 142 | 192 | 239 | 286 | 334 | 384 | 433 | 637 | 830 | 1023 | 1201 |
| 9M Opt | 33 | 74 | 120 | 173 | 238 | 295 | 354 | 414 | 473 | 529 | 766 | 991 | 1214 | 1424 |
| 1Y Opt | 39 | 86 | 142 | 209 | 281 | 347 | 416 | 484 | 553 | 618 | 888 | 1150 | 1403 | 1635 |
| 18M Opt | 52 | 114 | 186 | 272 | 357 | 438 | 520 | 600 | 679 | 757 | 1077 | 1383 | 1676 | 1959 |
| 2Y Opt | 70 | 149 | 238 | 333 | 430 | 519 | 607 | 695 | 781 | 869 | 1236 | 1578 | 1908 | 2224 |
| 3Y Opt | 106 | 213 | 323 | 434 | 547 | 651 | 753 | 853 | 950 | 1049 | 1474 | 1875 | 2266 | 2633 |
| 4Y Opt | 132 | 257 | 381 | 507 | 630 | 745 | 859 | 969 | 1075 | 1181 | 1639 | 2078 | 2507 | 2903 |
| 5Y Opt | 148 | 286 | 420 | 553 | 684 | 810 | 932 | 1051 | 1168 | 1279 | 1753 | 2212 | 2659 | 3079 |
| 7Y Opt | 166 | 317 | 462 | 606 | 747 | 883 | 1019 | 1152 | 1280 | 1406 | 1896 | 2361 | 2811 | 3238 |
| 10Y Opt | 172 | 333 | 485 | 635 | 783 | 930 | 1074 | 1215 | 1352 | 1486 | 1987 | 2450 | 2885 | 3288 |
| 15Y Opt | 171 | 328 | 481 | 631 | 778 | 922 | 1064 | 1205 | 1342 | 1486 | 1963 | 2399 | 2768 | 3132 |
| 20Y Opt | 162 | 313 | 464 | 611 | 756 | 899 | 1036 | 1172 | 1303 | 1429 | 1887 | 2275 | 2663 | 2963 |
| 25Y Opt | 156 | 300 | 445 | 581 | 716 | 847 | 973 | 1099 | 1222 | 1369 | 1793 | 2144 | 2492 | 2786 |
| 30Y Opt | 151 | 277 | 406 | 531 | 673 | 781 | 910 | 1038 | 1168 | 1304 | 1701 | 2038 | 2360 | 2659 |

**C** 31/05/2012 - EUR ATM Swaption Straddles - Fwd Premium Mids

| Option Expiry | 1Y | 2Y | 3Y | 4Y | 5Y | 6Y | 7Y | 8Y | 9Y | 10Y | 15Y | 20Y | 25Y | 30Y |
|---|---|---|---|---|---|---|---|---|---|---|---|---|---|---|
| 1M Opt | 12 | 24.5 | 39 | 56 | 73.5 | 92 | 112 | 133 | 154 | 177 | 271 | 362 | 445 | 523 |
| 2M Opt | 16.5 | 34.5 | 56 | 78.5 | 106 | 132 | 161 | 189 | 218 | 249 | 380 | 510 | 628 | 734 |
| 3M Opt | 19.5 | 43 | 69 | 100 | 131 | 164 | 198 | 234 | 271 | 310 | 461 | 613 | 753 | 877 |
| 6M Opt | 27.5 | 60 | 98.5 | 143 | 192 | 240 | 288 | 336 | 386 | 435 | 640 | 834 | 1027 | 1206 |
| 9M Opt | 33 | 75 | 121 | 175 | 239 | 297 | 356 | 416 | 477 | 534 | 772 | 999 | 1221 | 1434 |
| 1Y Opt | 39 | 86.5 | 143 | 211 | 283 | 350 | 419 | 487 | 557 | 622 | 894 | 1159 | 1412 | 1647 |
| 18M Opt | 52.5 | 115 | 188 | 274 | 360 | 442 | 525 | 606 | 687 | 765 | 1088 | 1398 | 1693 | 1981 |
| 2Y Opt | 71 | 151 | 241 | 337 | 435 | 526 | 615 | 704 | 792 | 882 | 1254 | 1602 | 1935 | 2258 |
| 3Y Opt | 107.5 | 217 | 329 | 443 | 560 | 666 | 770 | 873 | 973 | 1074 | 1509 | 1921 | 2319 | 2700 |
| 4Y Opt | 135 | 265 | 394 | 525 | 652 | 772 | 890 | 1004 | 1115 | 1224 | 1701 | 2157 | 2601 | 3019 |
| 5Y Opt | 154.5 | 301 | 443 | 583 | 721 | 854 | 983 | 1109 | 1231 | 1349 | 1854 | 2334 | 2806 | 3250 |
| 7Y Opt | 180 | 348 | 507 | 665 | 821 | 971 | 1119 | 1263 | 1405 | 1544 | 2086 | 2595 | 3094 | 3569 |
| 10Y Opt | 200 | 392 | 573 | 748 | 922 | 1094 | 1264 | 1431 | 1594 | 1753 | 2343 | 2889 | 3404 | 3880 |
| 15Y Opt | 221 | 437 | 641 | 844 | 1041 | 1234 | 1424 | 1611 | 1794 | 1983 | 2620 | 3203 | 3695 | 4182 |
| 20Y Opt | 234 | 463 | 684 | 900 | 1111 | 1318 | 1521 | 1720 | 1915 | 2102 | 2778 | 3348 | 3885 | 4362 |
| 25Y Opt | 241 | 478 | 710 | 932 | 1149 | 1363 | 1574 | 1783 | 1988 | 2197 | 2875 | 3440 | 3997 | 4467 |
| 30Y Opt | 251.5 | 494 | 727 | 951 | 1169 | 1386 | 1602 | 1816 | 2033 | 2262 | 2950 | 3534 | 4091 | 4612 |

**D** 31/05/2012 - EUR ATM Swaption Straddles - Implied Volatilities

| Option Expiry | 1Y | 2Y | 3Y | 4Y | 5Y | 6Y | 7Y | 8Y | 9Y | 10Y | 15Y | 20Y | 25Y | 30Y |
|---|---|---|---|---|---|---|---|---|---|---|---|---|---|---|
| 1M Opt | 92.6 | 65.3 | 63.9 | 61.8 | 57.9 | 54.7 | 53 | 51.9 | 51.4 | 51.2 | 48.6 | 50.7 | 52.9 | 54.9 |
| 2M Opt | 89.9 | 63.3 | 63.2 | 59.2 | 56.6 | 53.6 | 52 | 50.5 | 49.7 | 49.4 | 46.9 | 49.3 | 51.4 | 53.2 |
| 3M Opt | 89.8 | 64.4 | 62.8 | 60.7 | 56.5 | 53.4 | 51.5 | 50.5 | 49.9 | 49.7 | 46.3 | 48.1 | 50.2 | 51.7 |
| 6M Opt | 88.4 | 62.3 | 61.1 | 58.7 | 56.4 | 53.7 | 51.7 | 50.2 | 49.3 | 48.6 | 45.1 | 46.3 | 48.5 | 50.4 |
| 9M Opt | 85.9 | 61.8 | 58.9 | 56.2 | 55.1 | 52.6 | 50.9 | 49.8 | 48.9 | 47.9 | 44.2 | 45.2 | 47.1 | 48.9 |
| 1Y Opt | 84.4 | 59.4 | 57.7 | 55.9 | 54.2 | 51.9 | 50.4 | 49.2 | 48.4 | 47.4 | 44 | 45.2 | 47 | 48.6 |
| 18M Opt | 80.8 | 58.2 | 55.2 | 53.9 | 52.1 | 50.2 | 48.8 | 47.6 | 46.7 | 45.9 | 43 | 44.2 | 45.9 | 47.5 |
| 2Y Opt | 81.5 | 58.9 | 55.1 | 52.4 | 50.7 | 48.6 | 47 | 45.8 | 44.8 | 44.2 | 42.3 | 43.5 | 45.2 | 46.8 |
| 3Y Opt | 72.6 | 54.8 | 50.6 | 48.3 | 47 | 45.2 | 43.9 | 42.8 | 41.9 | 41.5 | 40.6 | 42.1 | 44 | 45.4 |
| 4Y Opt | 58.5 | 48.1 | 45.7 | 44.4 | 43.2 | 41.9 | 41 | 40.1 | 39.5 | 39.3 | 39.1 | 40.7 | 42.6 | 43.8 |
| 5Y Opt | 50 | 44 | 42.3 | 41.1 | 40.2 | 39.4 | 38.7 | 38.2 | 37.9 | 38 | 39.5 | 41.3 | 42.3 |
| 7Y Opt | 42.5 | 38.8 | 37.4 | 36.6 | 36.1 | 35.8 | 35.8 | 35.9 | 36.2 | 36.6 | 36.9 | 38 | 39.2 | 39.6 |
| 10Y Opt | 35.1 | 33.6 | 33.2 | 33.2 | 33.6 | 34.1 | 34.9 | 35.7 | 36.5 | 37.1 | 37.1 | 37.6 | 37.6 | 37.1 |
| 15Y Opt | 37 | 37.2 | 37.9 | 38.8 | 39.6 | 40.2 | 40.8 | 41.3 | 41.8 | 42.6 | 40.3 | 38.2 | 35.7 | 34.6 |
| 20Y Opt | 44.9 | 43.8 | 43.9 | 44.1 | 44 | 44.9 | 45.4 | 45.7 | 45.9 | 45.7 | 39.6 | 35.1 | 33.1 | 31.7 |
| 25Y Opt | 49.1 | 48.1 | 48.3 | 47.6 | 46.7 | 45.7 | 44.6 | 43.6 | 42.7 | 43 | 35 | 31.4 | 29.8 | 28.6 |
| 30Y Opt | 45.7 | 40.2 | 38.6 | 37.2 | 37.6 | 35.5 | 35.1 | 34.8 | 34.8 | 35.1 | 30.5 | 28 | 26.8 | 26.3 |

**Figure 19:** EUR at-the-money Swaption market quotes on 31 May 2012. Premia on the left upper panel (panel A) are spot premia obtained from forward premia (left lower panel, panel C) using Eonia discounting. Premia on the right upper panel (panel B) are spot premia obtained by discounting forward premia using the Euribor yield curve. There is an unique ATM implied volatility surface (right lower panel, panel D), consistent with multiple-curve CSA discounting methodology. (Source: ICAP).

## 3.5. Market Practice and P&L Impacts

Up to the end of 2010, just a few banks and clearing houses have declared full adoption of CSA discounting also for balance sheet revaluation and collateral margination (see e.g. Bianchetti 2012). On 17 June 2010 LCH.Clearnet communicated that its clearing platform SwapClear switched to OIS discounting for its $218 trillion Interest Rate Swap portfolio, in line with the new market practice for collateralized trades (Whittall 2010). The ISDA Margin Survey 2012 reports information regarding the diffusion of OIS and CSA discounting. ISDA distinguishes between OIS discounting, based on the use of an OIS yield curve for discounting purposes, and CSA discounting, based on the intent of reflecting implied economic terms within the deal valuation process. Data reported in Table 3 represent the percentage of 12 respondents that affirm to price at least a subset of OTC derivatives for margin purposes adopting OIS or CSA discounting.

|  | OIS Discounting | CSA Discounting |
|---|---|---|
| Commodity Derivatives | 16.6% | 25.0% |
| Credit Derivatives | 33.3% | 33.3% |
| Equity Derivatives | 25.0% | 33.3% |
| Fixed Income Derivatives | 58.3% | 50.0% |
| FX Derivatives | 16.6% | 33.3% |

**Table 3**: percentage of 12 respondents to the ISDA Margin Survey 2012 pricing at least some OTC derivatives using OIS or CSA discounting (source: ISDA 2012).

The embracing of the CSA discounting can determine relevant balance sheet impacts when a financial institution is re-valuing its portfolio considering the funding implications embedded in the collateral agreement. During the year 2010 some banks reported the NPV variations experienced on their OTC derivatives portfolios due to the adoption of the CSA discounting methodology. For example, BNP Paribas has declared € 108 mln loss on its IRS portfolio, instead Morgan Stanley has stated $ 176 mln gain from its IRD positions, Credit Agricole has accounted a negative variation on its Fixed Income portfolio of € 120 mln, while Royal Bank of Scotland and UBS has communicated a profit of £ 127 mln and CHF 76 mln respectively (see Cameron 2011 for further details). Clearly, the size and the direction of the P&L impacts are strongly influenced by the composition and the structure of the portfolio involved in the revision of the discounting methodology (Cameron 2011).

## 3.6. Issues of CSA Discounting

Switching financial institutions to CSA discounting in practice is not an easy task at all because of a variety of issues, that we discuss in the subsections below.

### 3.6.1. Collateral and Liquidity Issues

Besides the Evidence of CSA discounting from collateral management may be controversial. Collateral margination is usually managed by collateral desks at portfolio level for each counterparty under CSA, and not at trade level, thus hiding the discounting effects. On the other hand, in case of disputation pricing details on single or few trades are shared between the two counterparties in order to match the mark to market, thus allowing much more market intelligence than usual. Complications may arise because of the typical variety of clauses and details of collateral agreements, such as haircuts, margination frequency, rate spreads, currency, one-way margination, etc. that require, in principle, more sophisticated and CSA dependent pricing methodologies. Another bias may be introduced by opportunistic counterparties posting or asking collateral using their most convenient discounting methodology. The ISDA Standard CSA discussed in section 3.3 addresses and simplifies these issues.

A very important challenge is the front-to-back integration of Banks' internal credit and funding management, from trading to treasury, collateral and back office, in order to benefit of centralised credit and liquidity charges at single trade level. Such a re-organisation of traditionally separated areas may result to be very difficult, in particular for global international banking groups characterised by multiple subsidiaries and locations. In particular, the yield curves used for pricing internal deals (trades between different legal entities inside the banking group) reflects the cost of internal funding within the group, and has to do with the transfer pricing policy and business model of the Bank.

### 3.6.2. Accounting Issues

Since the International Accounting Standards (IAS), issued by the International Accounting Standards Board (IASB), stating that "in determining the valuation of OTC derivative [...] a valuation technique (a) incorporates all factors that market participants would consider in setting a price and (b) is consistent with accepted economic methodologies for pricing financial instruments" (AG76), there exist a judgemental area, where the estimation of fair value is based on market (multilateral) consensus. CSA discounting is a typical case of evolving market consensus regarding the nature of CSA, from a simple accessory legal guarantee to a determinant of the fair value.

Hedge accounting, in particular, is an accountancy practice allowed by IAS to mitigate the Profit & Loss volatility due to derivatives used for hedging. A typical situation arises when the interest rate risk of a liability (a bond issued by the bank, for instance) is hedged using a Swap. Hedge accounting requires that the profit & loss of the package (Bond + Swap) remains confined in the 80%-125% window with respect to the initial fair value. The pricing of the package is based on ad hoc methodology (e.g. the liability cash flows are discounted using the floating rate of the Swap, for instance), that may partially accounts for the basis risk existing between the liability and the derivative. As a consequence the adoption of CSA discounting may realize the basis risk, resulting in significant NPV jumps and even breaches of the hedge accounting 80-125 constrain. Hence, either the methodology must be revised to account for the basis risk, or hedges must be renegotiated.

In order to converge with the principles issued by the Financial Accounting Standards Board (FASB) prescribed by the FAS157 (FASB 2006), the IASB has issued a new International Financial Reporting Standard (IFRS), in force since January 2013. According to the IFRS13, "*the fair value is defined as the price that would be received to sell an asset or paid to transfer a liability in an orderly transaction between market participants at the measurement date*" (IASB 2012).

The determination of the fair value is, hence, a market-based measurement, not an entity-specific measurement. When measuring fair value, an entity uses the assumptions that market participants would use when pricing the asset or liability under current market conditions, including assumptions about risk. As a result, an entity's intention to hold an asset or to settle or otherwise fulfil a liability is not relevant when measuring fair value. It is an exit price at the measurement date from the perspective of a market participant that holds the asset or owes the liability. In the case of OTC collateralized deals, the fair value of the contract must consider the effect of collateralization and, thus, it has to be consistent with the adoption of the CSA discounting approach. Moreover, the IFRS13, as the FAS157, allows to include in the fair value of an asset or liability the adjustments related both to the counterparty's credit risk and to the entity's own credit risk (Credit/Debit Valuation Adjustment, CVA/DVA, respectively).

Responding to the market evolution and facing with industry needs, the FASB considers to introduce the use of the federal funds rate as the benchmark rate in the fair value determination of collateralized trades in US dollars (Madigan 2012).

### 3.6.3. IT Issues

The adoption of CSA-discounting is a big issue from an IT point of view that requires huge resources to be properly addressed. Here are some critical points.

- Booking of trades in pricing systems must be reviewed, such that the information regarding the associated collateral is recovered.
- Multiple yield curves and volatilities bootstrapping must be properly and consistently configured across all pricing systems.
- Pricing systems configurations must be reviewed for CSA compliance, allowing proper assignments to each trade of different yield curves depending on the CSA. Hidden assumptions regarding discounting, e.g. default assignments of yield curve usage without explicit flags must be carefully avoided.
- Risk computations and systems must be reviewed as well, to capture the effects of the larger set of risk factors implied by multiple-curve CSA discounting methodology.
- Commercial systems require new releases able to manage CSA discounting. Vendors must be typically fed with appropriate specs and the new releases carefully tested.
- Proprietary systems and financial libraries must be reviewed and re-engineered to make them multiple-curve compliant. Previous poor library design is likely to require much more re-implementation effort.
- Systems integration and alignment must be carefully checked to avoid the classical "two systems two prices" problem.

In general, we can say that the switch to CSA discounting is a stress test for the IT architecture of a bank. The most complex or confused IT situations typically imply much more effort to switch, and vice versa.

### 3.6.4. Risk Management Issues

The adoption of CSA-discounting is a big issue also from a risk management point of view. We discuss each kind of risk in the points below.

- **Model risk**: this source of risk has to do with the modelling choices adopted for pricing trades and computing the corresponding risk measures. From this point of view, model risk is a primary source of risk underlying all the classical risk management areas discussed below (market risk, credit risk, etc.). The first and most important model risk in CSA discounting is to rely on the classical framework for pricing derivatives, in particular in presence of large basis expositions. In particular the market standard for uncollateralised trades is still under development for what regards the inclusion of the funding spread, leading to a so-called Funding Valuation Adjustment (FVA, see e.g. Carver 2012).
  A second source of model risk may be hidden into the clauses of collateral agreements, such as multiple eligible collateral assets and currencies, initial margin, close outs, etc. Once these details are included in the pricing methodology, important NPV jumps may appear, depending on the exposure of the bank with respect to it's market counterparties. Finally, another source of model risk are the modern multiple-curve pricing models. Even if these models may be able to give a better description of the basis risk, they are, to date, still under development, and there is neither market standard nor quotations available for complete calibration, such as OIS options and volatilities.

- **Market risk**: the most important source of market risk involved in CSA discounting is the basis risk in the multiple-curve world, in which even plain vanilla interest rate derivatives (e.g. Swaps) display complex delta sensitivities and exposures distributed across multiple Libors with different tenors and OIS rates. This kind of risk may be not fully captured or represented in standard, old style pricing and risk management frameworks grounded on Libor discounting. Basis risk is also expensive to hedge, requiring market Swaps, OIS and Basis Swaps. Furthermore, hedging the basis risk volatility would require options on the basis, not presently quoted in the market. In practice, basis risk is often hedged by proxy, using standard Libor Swaps and the most liquid Basis Swaps, thus leaving an open exposure to the Libor-OIS basis. The latter may be huge (Figure 7) and volatile (Figure 1). The corresponding (un)expected profit & loss is typically realized in case of unwindings or in case of adoption of CSA discounting, for instance when trades are migrated to Central Counterparties.

- **Credit and counterparty risk**: this source of risk is captured in CSA discounting in the sense that, for trades under CSA, the collateral reduces the counterparty risk and the CSA discounting ensures no-arbitrage between the collateral rate and the discounting rate. A residual source of counterparty risk is left behind by re-hypothecation issues and by the mechanics of margination (see Brigo et al. 2011). In case of absence of CSA, Credit Value Adjustment (CVA) and Debt Value Adjustment (DVA) must be calculated, according with the funding component (FVA). We stress that a consistent treatment of DVA and FVA is an open topic still under investigation (see e.g. Morini and Prampolini 2011, Fries 2010, Carver 2012).

- **Liquidity and funding risk**: with liquidity and funding risk we mean the risk induced by the volatility of market funding rates. Funding liquidity risk management under CSA discounting is complicated by the fact that derivatives have a funding impact that depends on the CSA. The situation for uncollateralised trades is even more complex, because of the unclear funding component of the fair value (FVA) and of the uncertain and complex nature of the funding curve, depending on the prevailing market funding channels of the bank. In any case, a centralised liquidity management, integrating treasury, collateral management and sales/trading desks, would allow both a full view of all the expected cash flows generated by the bank's activity by derivatives in particular, and a correct pricing of funding costs at single trade level.

- **Operational risk**: the main source of operational risk (the risk of loss resulting from failed internal processes, people, systems, or external events) generated by CSA discounting is related to the increasing complication of pricing systems and liquidity management discussed above. A typical example may be a wrong assignment between a deal or a group of deals and their CSA, resulting in a wrong pricing. An unexpected Profit & Loss is revealed when the mistake is fixed.

We conclude with the observation that the main driver of the switch to CSA discounting is the evolution of pricing and risk methodologies, under the pressure of market evolution after the credit crunch. This a typical situation in which a solid Risk Management with strong quantitative resources may serve both as the traditional defence against unexpected losses, and as the pivot of the innovation.

### 3.6.5. Management Issues

Management is called to lead the change, and the corresponding frictions, taking business opportunities and controlling risks and costs. The main management decisions required for switching to CSA-discounting, as discussed in the points above, regard:

- timing: when to switch
- how to switch: all together or piecewise, depending on currency, asset classes, desks, subsidiaries, time-zone, main trading markets, etc.
- a clear view about the multiple funding sources of the Bank (the funding curve) and re-organisation for centralised credit and liquidity management
- review and cleaning of collateral agreements with counterparties
- how to manage the basis risk and the Profit & Loss generated by the switch
- how to manage the hedge accounting
- IT upgrade: booking, pricing, reporting, etc.
- communication and explanation of the switch to markets, customers, auditors and regulators.

### 3.6.6. The Role of Quants

It is clear from the discussion above that CSA discounting is a typical complex problem in which a simple no-arbitrage pricing issue (choosing the correct discounting curve) generates many consequences that propagate all around in the market and inside the banks. In such a situation quant people have the responsibility of extending the modern no-arbitrage pricing framework into other areas of the bank, traditionally not familiar with pricing issues, in order to reach a better fair value and risk management at Bank's level. Citing the conclusion of the KPMG survey (KPMG 2011), "CSA or funding related valuation is not a pure playground for quants, but rather a topic that evokes questions about transfer pricing, steering of risk, and, most importantly, the business model of each bank."

# 4. Conclusions

In this work we have presented a qualitative analysis of the markets evolution after the begun of the financial crisis in 2007. In particular, we have focused on the fixed income market and we have reported the most relevant empirical evidences regarding the divergences between the Euribor and Eonia OIS rates, between FRA and forward rates and the explosion of Basis Swap spreads. These market frictions have induced a segmentation of the interest rate market into sub-areas, corresponding to instruments with risky underlying Euribor rates distinct by tenors, and almost risk free overnight rates, characterized, in principle, by different internal dynamics reflecting different credit and liquidity risks.

In response to the crisis, the classical pricing framework, based on a single yield curve used to calculate forward rates and discount factors, has been abandoned, and a new modern pricing approach has prevailed among practitioners, taking into account the market segmentation as an empirical evidence and incorporating the new interest rate dynamics into a multiple curve framework. We have shown market evidences of the diffusion and of the mechanism of collateral agreements among interbank dealers since the beginning of the financial crisis. Next, we have introduced the multiple curve pricing framework, called CSA discounting, in the evaluation of collateralized contracts, showing that under no-arbitrage and self-financing assumptions the discounting curve must reflects the funding rate of the contract, that, in the case of collateralized OTC derivatives, usually coincides with the relevant O/N interest rate (i.e. Eonia for the EUR market). Consequently, we have reported evidences of the market transition to the modern CSA discounting pricing approach, and discussed the most relevant issues that a financial institution has to consider.

Across all the paper, we argument that the roots of the numerous and complex changes encountered on the market in these years can be found in the different credit and liquidity risk perception of the interbank market participants, that can no longer consider themselves as "too big to fail". We also believe that such risks and the corresponding consequences, such as the Libor-OIS basis, will not return negligible as in the pre-crisis world, and will be there in future, exactly as the volatility smile has been there since the 1987 market crash.

Expected further developments will regard, for example, the investigation of the relevant risk factors reflected in Libor rates (see e.g. Filipovic and Trolle 2012) and the pricing of non-collateralized derivatives considering the bilateral default risk of the counterparties in terms of Credit Value Adjustment (CVA) and Debt Value Adjustment (DVA) and liquidity risk in the form of Funding Value Adjustment (FVA) (see Morini and Prampolini 2011).